\documentclass[amsmath,amssymb,pre,twocolumn,a4paper]{revtex4}

\usepackage{inconsolata}
\usepackage[bbgreekl]{mathbbol} 
\usepackage{amsfonts,amsthm,amsmath,amssymb,mathrsfs,mathtools}
\usepackage{dsfont}
\usepackage{epsfig, color} 
\usepackage{verbatim,subfigure,hyperref,framed,graphics,graphicx,wrapfig,textcomp}
\usepackage{epstopdf}

\newcommand{\gML}[4]{E^{#3}_{#1,#2}\left( #4 \right)}

\newcommand*{\diff}[1]{\mathop{}\!\mathrm{d} #1  }
\newcommand{\derpar}[2]{\frac{\partial #1}{\partial #2}}
\newcommand{\dersecpar}[2]{\frac{\partial^2 #1}{\partial #2^2}}
\newcommand{\dertot}[2]{\frac{\diff #1}{\diff #2}}

\newcommand{\Average}[1]{\left< #1 \right>}

\begin{document}

\title{Weak Galilean invariance as a selection principle for coarse-grained diffusive models}

\author{Andrea Cairoli$^{1,2}$, Rainer Klages$^2$, and Adrian Baule$^{2,}$\footnote{To whom correspondence should be addressed.\\ Email: a.baule@qmul.ac.uk.}}
\affiliation{$^1$Department of Bioengineering, Imperial College London, London SW7 2AZ, UK\\
$^2$School of Mathematical Sciences, Queen Mary University of London, London E1 4NS, UK}

\begin{abstract}
  How does the mathematical description of a system change in
  different reference frames? Galilei first addressed this
  fundamental question by formulating the famous principle of Galilean
  invariance. It prescribes that the equations of motion of closed
  systems remain the same in different inertial frames related by
  Galilean transformations, thus imposing strong
  constraints on the dynamical rules. However, real world systems are
  often described by coarse-grained models integrating complex
  internal and external interactions indistinguishably as friction and stochastic
  forces. Since Galilean invariance is then violated, there is seemingly no
  alternative principle to assess a priori the physical consistency
  of a given stochastic model in different inertial frames. Here, starting from the Kac-Zwanzig
  Hamiltonian model generating Brownian motion, we show how Galilean
  invariance is broken during the coarse graining procedure when
  deriving stochastic equations. Our analysis leads to a set of rules
  characterizing systems in different inertial frames that have to be
  satisfied by general stochastic models, which we call ``weak
  Galilean invariance". Several well-known stochastic processes are
  invariant in these terms, except the continuous-time random walk for
  which we derive the correct invariant description. Our results are
  particularly relevant for the modelling of biological systems, as
  they provide a theoretical principle to select physically consistent stochastic models prior to a validation against experimental data.
  
\end{abstract}

\maketitle 


%


\begin{figure*}[!htb]
\centering
\includegraphics[width=160mm]{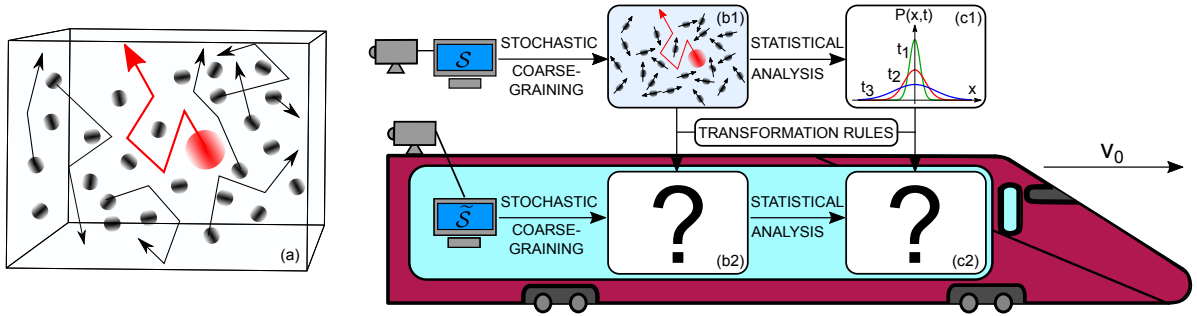}
\caption{
Pictorial representation of the setup: a system of $N$
    heat bath particles (black) and one tracer (red) is observed from
    two different reference frames $\mathcal{S}$ and
    $\widetilde{\mathcal{S}}$. While $\mathcal{S}$ is at rest,
  $\widetilde{\mathcal{S}}$ is moving with velocity $v_0$ with respect
  to $\mathcal{S}$. We consider three different levels of description
  of the original system: (a) The microscopic system of $N+1$
  particles is described by deterministic equations of motion leading
  to trajectories fully specified by the initial conditions. (b1)
  Alternatively, one can provide a stochastic coarse-grained model of
  the tracer dynamics in terms of effective dissipative friction
  forces and random collisions with the $N$ bath particles (arrowed
  spheres), which account for their original microscopic interactions
  with the probe. (c1) Finally, the system can be studied in terms of
  its position and velocity statistics, whose distributions are
  uniquely determined by either experimental measurements or by a
  prescribed stochastic model. While the relationship between the
    dynamical evolutions in $\mathcal{S}$ and $\tilde{\mathcal{S}}$
    for (a) is specified by Galilean
    transformations of the position and velocity degrees of freedoms,
    here we derive the corresponding relationships 
    for (b1) $\leftrightarrow$ (b2) and (c1) $\leftrightarrow$ (c2) yielding what we call 
    {\em weak Galilean invariance}.
}\label{cartoon}
\end{figure*}

Classical mechanics is built upon the two intimately related concepts
of inertial reference frames and Galilean invariance (GI)
\cite{arnol2013mathematical}. The former are coordinate systems where
a freely moving particle (i.e., in the absence of external forces) either is at rest or exhibits
uniform rectilinear motion. The latter principle states that in
different inertial frames the equations of motion of closed systems,
i.e., including all their interacting constituents, are invariant with
respect to Galilean transformations (GTs). 
These are in general affine transformations, that preserve both time intervals and distances between simultaneous events \cite{arnol2013mathematical}. For systems whose dynamical evolution can be fully characterized by microscopic deterministic models, GI plays a fundamental constitutive role, manifest in the constraints that it naturally imposes on the functional form of Newton's equation. However, a large variety of complex systems in science and nature are not modelled on a microscopic level with Newtonian equations of motion, but rather on a {\it mesoscopic level} using, e.g., stochastic Langevin equations or Fokker-Planck diffusion equations to capture the coarse-grained effects of microscopic interactions as friction and noise on the relevant degrees of freedom. The applications of such equations and their variants are vast throughout the sciences \cite{Van-Kampen:2011aa,Gardiner:2010aa,Kalmykov:2012aa}.

Coarse-grained diffusive models are particularly relevant to describe anomalous transport phenomena, where stochasticity arises due to complex multi-particle interactions, whose precise form is usually unknown. While for normal diffusion due to Brownian motion the mean-square displacement (MSD) of an ensemble of particles with positions $X(t)$ at time $t$ grows linearly in the long-time limit, $\langle X^2\rangle\sim t^{\beta}$ with $\beta=1$, for anomalous diffusion it scales non-linearly with $\beta\neq1$. Anomalous {dynamics} has been observed experimentally for a wide range of physical processes like 
{particle transport in plasmas, molecular diffusion in nanopores and charge transport in amorphous semiconductors \cite{metzler2000random,mekl04,klages2008anomalous}, 
that was first theoretically described in \cite{scher1973stochastic,scher1975anomalous} based on the Continuous time random walk (CTRW) \cite{montroll1965random}.} 
{Likewise, anomalous diffusion has been later found for biological motion \cite{HoFr13,Dieterich:2008aa,Harris2012}, and even human movement \cite{Brockmann:2006aa}.}  Recently, it has been established as an ubiquitous characteristic of cellular processes on a molecular level \cite{Bressloff:2013aa}. 
Here, anomalous diffusion is observed, e.g., in neuronal messenger ribonucleoprotein transport \cite{Song2018}, in protein structural fluctuations \cite{Hu:2015aa}, and in the intracellular transport of S. cerevisiae mitochondria \cite{senning2010actin}, chromosomal loci of E.~coli cells \cite{Weber2010,Javer2014}, engulfed microspheres \cite{Caspi2000}, lipid and insulin granules \cite{Jeon2011,tabei2013intracellular}. However, because of the intrinsic difficulties in assessing the details of the microscopic interactions in experiments, theoretical models for such anomalous processes cannot be typically derived from first principles and are usually formulated on mostly phenomenological grounds. In fact, a wealth of diffusive models has been suggested in the literature, which rely on spatiotemporal memory effects and non-Gaussian power-law statistics of various observables \cite{metzler2000random,klages2008anomalous,zaburdaev2015levy,Meroz:2015aa}. Unfortunately so far there is
no fundamental rule available that could be employed to verify the physical consistency of such stochastic models {\em a priori}. To distinguish between different models it remains only the comparison with experimental data that is often imprecise due to limited sample sizes.

Here, we show that GI can provide precisely such a constitutive principle. Even though the fundamental role of GI seemingly breaks down for stochastic diffusive models due to the presence of friction \cite{sekimoto2010stochastic}, they are nevertheless constrained by a weak form of GI in order to be physically consistent in different inertial frames. The weak GI rules derived below thus represent a general selection principle for stochastic coarse-grained models. Previously, the consequences of GI in the context of statistical mechanics were first explored for fluid dynamics,
where it establishes specific relations between critical exponents of
the characteristic parameters entering the derivation of the
Navier-Stokes equation \cite{forster1977large} (although this result 
has been challenged \cite{berera2007gauge}). 
The problem carries over to the famous
KPZ equation \cite{kardar1986dynamic} whose GI is equally debated \cite{wio2010kpz}. 
Whether or not these statistical equations feature
GI has important practical implications for the modelling of, e.g., fluid flows
\cite{berera2007gauge} and nonlinear biological growth
\cite{escudero2010}. Specifically, 
in molecular dynamics simulations of fluids employing stochastic Langevin
thermostats it was found that Langevin dynamics breaks GI by violating
global momentum conservation, which makes it unsuitable to simulate hydrodynamic phenomena \cite{duenweg1993}. Curing this
deficiency led to novel GI algorithms, most notably dissipative
particle dynamics, now widely used to simulate soft matter systems and
simple liquids \cite{hoogerbrugge1992simulating,soddemann2003,pastorino2015}.

The basic setup of our problem is represented in
Fig.~\ref{cartoon}: here $\mathcal{S}$ and $\widetilde{\mathcal{S}}$
are two inertial reference frames, where $\mathcal{S}$ is the
laboratory frame at rest while $\widetilde{\mathcal{S}}$ is moving
with uniform velocity $v_0$ with respect to $\mathcal{S}$. The GTs
connecting the coordinates in the two frames are given by
\begin{equation}
\widetilde{x}=x-v_0 t\:, \qquad\qquad \widetilde{v}=v-v_0\:, \label{eq:Gtransf} 
\end{equation}
where, for simplicity, we focus on the one dimensional case. \eqref{eq:Gtransf} is the phase space version of the classical GTs assuming an absolute time \cite{arnol2013mathematical}. A
classical system of $N+1$ interacting particles is described by the
Hamiltonian function
\begin{multline}
\label{Hamiltonian}
H(x_1,v_1;\ldots ;x_{N+1},v_{N+1})=\\ \sum_{i} \frac{m_i}{2} v_i^2(t) + \sum_{i<j} U(x_i(t),x_j(t)) \:,
\end{multline}
where $x_i$, $v_i$ are the position-velocity coordinates of the $i$-th
particle in the reference frame $\mathcal{S}$ and $U$ is the
interaction potential satisfying some mild regularity conditions.  Its
dynamics is specified by Hamilton's equations
\begin{equation}
\label{HamiltonEOM}
\dot{x}_i(t)=v_i(t)\:,\quad
m_i\dot{v}_i(t)=-\derpar{}{x_i} \sum_{i<j} U(x_i(t),x_j(t))\:.
\end{equation}  
Transforming the coordinates to the reference frame
$\widetilde{\mathcal{S}}$ via Eqs.~(\ref{eq:Gtransf}), we see that
$\dot{\widetilde{x}}_i(t)=\widetilde{v}_i(t)$ and
$m_i\dot{\widetilde{v}}_i(t)=-\derpar{}{\widetilde{x}_i}
\sum_{i<j} U(\widetilde{x}_i(t),\widetilde{x}_j(t))$ if $U$ depends
only on the relative difference between the particles' positions, i.e., $U(x_i(t),x_j(t))=U(x_i(t)-x_j(t))$,
because in this case $\widetilde{x}_i(t)-\widetilde{x}_j(t)=x_i(t)-x_j(t)$. We
thus recover Newton's equations of motion satisfying his Third Law,
which are identical in both reference frames, i.e., they satisfy
GI. Our goal is now to derive coarse-grained dynamics from systems
described by Eqs.~(\ref{HamiltonEOM}), where some of the microscopic
degrees of freedom have been eliminated, and to characterize their
statistics on such a mesoscopic level in both frames $\mathcal{S}$, $\widetilde{\mathcal{S}}$
(see Fig.~\ref{cartoon}).

The transition from
Eqs.~(\ref{HamiltonEOM}) to an effective description in the form of a stochastic diffusion equation can be made quantitatively precise for the specific scenario where 
one of the particles, for simplicity let it be the $(N+1)$-th, is a tagged (tracer) particle of mass $m_{N+1}=M$, that interacts with the remaining particles of equal mass $m_j=m$ via an harmonic potential of coupling strength $m \omega^2_j$, thus defining the environment as a heat bath, i.e., 
$U(X,x_j)\to \sum_{j=1}^N m\omega_j^2[X(t)-x_j(t)]^2/2$. 
Conversely, interactions between different bath particles are switched off.
This is a Galilean invariant version of the classical Kac-Zwanzig model \cite{zwanzig1973nonlinear}, whose relevance has been recently addressed \cite{deBacco2014}. Denoting by $(X(t),V(t))$ and $(x_j(t),v_j(t)),\, j\!=\!1,...,N$
the position and velocity variables of the tracer and heat bath particles, respectively, in the frame
$\mathcal{S}$, their Hamilton's equations 
become: $M \ddot{X}(t)=\sum_{j=1}^N m \omega_j^2 \left[x_j(t)-X(t)\right]$ and $m \ddot{x}_j(t)= -m \omega_j^2 \left[x_j(t)-X(t)\right]$.
These equations specify the time evolution of all $N+1$ particles of
the system (arrows in the box of Fig.~\ref{cartoon}a) in $\mathcal{S}$
once the initial conditions are prescribed, which we take as
$(X(0),V(0))=(0,0)$ and $(x_j(0),v_j(0))=(x_{j 0},v_{j 0})$,
without loss of generality. 
The great advantage of this model is that the effective dynamics for the tracer can be derived by integrating out the bath degrees of freedom. This yields \cite{zwanzig1973nonlinear}
\begin{equation}
\label{GLE}
M \ddot{X}(t)=-\int_0^t \Omega(t-t^{\prime}) \dot{X}(t^{\prime}) \diff{t^{\prime}} + \xi(t)\:,
\end{equation}  
where the memory kernel $\Omega$ and what later on will become the
``noise" $\xi$ in Langevin dynamics are exactly
\cite{zwanzig1973nonlinear}
\begin{align}
\label{kernel}
\Omega(t)&=\sum_{j=1}^N \omega_j^2 \cos{(\omega_j t)} \:, \\
\label{noise}
\xi(t)&=\sum_{j=1}^N  \omega_j v_{j 0} \sin{(\omega_j t)}+\sum_{j=1}^N \omega_j^2 x_{j 0} \cos{(\omega_j t)} \:.
\end{align}
As can be seen from \eqref{noise}, $\xi$ depends explicitly on the
initial conditions of the bath particles, which are related to those
in $\widetilde{\mathcal{S}}$ by $\widetilde{x}_{j 0}=x_{j 0}$,
$\widetilde{v}_{j 0}=v_{j 0}-v_0$ and $\widetilde{X}(0)=0$,
$\widetilde{V}(0)=-v_0$. Since everything is exact, the dynamics
in $\widetilde{\mathcal{S}}$ follows by applying the GTs of
Eqs.~(\ref{eq:Gtransf}) to Eqs.~(\ref{GLE}--\ref{noise}). $\Omega$ is
unchanged under the transformation, but \eqref{noise} is changed
due to the GTs of the initial velocities of the bath particles. If we
call $\widetilde{\xi}$ the noise term in the transformed frame, i.e.,
\eqref{noise} in $\sim$ variables, the two noises are related by
\begin{equation}
\xi(t)=\widetilde{\xi}(t)+v_0 \sum_{j=1}^N \omega_j \sin{(\omega_j t)}= \widetilde{\xi}(t)+v_0\int_0^t\Omega(t^{\prime})\diff{t^{\prime}}\:.
\label{eq:Frule}
\end{equation}
Overall, the deterministic coarse-grained equation of the tracer in
$\widetilde{\mathcal{S}}$ is then just \eqref{GLE} in $\sim$ variables
\begin{align}
  M \ddot{\widetilde{X}}(t)&=-\int_0^{t} \Omega(t-t^{\prime}) \dot{\widetilde{X}}(t^{\prime})\diff{t^{\prime}}+\widetilde{\xi}(t) \label{GLEboth}\\
  &= -\int_0^{t} \Omega(t-t^{\prime})
  (\dot{\widetilde{X}}(t^{\prime})+v_0)\diff{t^{\prime}}+\xi(t),
\label{GLEtilde}
\end{align}
using \eqref{eq:Frule}. The deterministic effective equation of motion for the tracer thus maintains the GI of the original microscopic dynamics even after projecting out the degrees of freedom of the bath particles. For deriving stochastic Langevin dynamics the next step is to
simplify this coarse-grained description by specifying $\xi(t)$ as a
random force instead of the deterministic force
\eqref{noise}. On the Langevin level, the dynamics of the
tracer then effectively originates from both dissipative friction
forces and random collisions with the bath particles, accounting for
their original microscopic interactions with the probe. The statistics
of $\xi(t)$ is specified by the distribution of $x_{j 0}$, $v_{j
  0}$. Assuming that the heat bath is at equilibrium in $\mathcal{S}$,
the velocity distribution is Maxwellian at the temperature of the
system $T$ implying $\left< \xi(t) \right>\!=\!0$ and $\left< \xi(t_1)
  \xi(t_2) \right>=k_{B} T \Omega(|t_1-t_2|)$
\cite{zwanzig1973nonlinear}. Consequently, the fluctuation-dissipation
relation holds \cite{kubo1966fluctuation}. Equation~(\ref{GLE}) then
defines a generalized Langevin equation (LE) in $\mathcal{S}$.
Crucially, the notion of thermal equilibrium is not frame invariant
such that the stochastic coarse-graining is not possible directly for
\eqref{GLEboth}. Specifying the properties of the random force
that way \textit{per se} singles out a reference frame and thus
inevitably breaks GI, because according to \eqref{eq:Frule}
  the noise $\tilde{\xi}$ acquires a different statistics than $\xi$.
However, after having specified $\xi$ via the equilibrium assumption
in $\mathcal{S}$, \eqref{GLEtilde} is still
valid. Eqs.~(\ref{GLE},\ref{GLEtilde}) then both represent the
same microscopic dynamics in two different inertial frames. We see
that \eqref{GLEtilde} contains an additional drift term, which
could be obtained directly from \eqref{GLE} by performing a GT on
the coordinates of its {\em deterministic part} 
only while leaving the noise term unchanged.

The transformation rules of the stochastic equations of motion imply
that the resulting position-velocity processes ($X$, $V$) and ($\widetilde{X}$, $\widetilde{V}$) are related via
a GT, even in the presence of stochasticity, which can be shown 
by explicitly solving these equations~(\ref{GLE},\ref{GLEtilde}), while correctly accounting for the different initial conditions in the two frames 
(Appendix \ref{rule}).
Consequently, also the probability density functions (PDFs) for
position and velocity in different inertial frames can be related to
each other directly. Including the position coordinates as
$\dot{X}(t)=V(t)$ and
$\skew{3.5}\dot{\widetilde{X}}(t)=\widetilde{V}(t)$ we have for
underdamped dynamics the PDF transformation rule
\begin{eqnarray}
P(x,v,t)&=&\left<\delta(x-X(t))\delta(v-V(t))\right>\nonumber\\
&=&\left<\delta(x-\widetilde{X}(t)-v_0t)\delta(v-\widetilde{V}(t)-v_0)\right>\nonumber\\
&=& \widetilde{P}(x-v_0t,v-v_0,t)\:,
\label{newtrans1}  
\end{eqnarray}
since the expected value in both inertial frames is over the
fluctuations of the same heat bath defined in $\mathcal{S}$. In terms
of its Fourier-Laplace transform 
(from now on denoted by different independent variables according to $(x,v,t)\to(k,p,\lambda)$) 
the connection is
$P(k,p,\lambda)=e^{-i p v_0}\widetilde{P}(k,p,\lambda -i k v_0)$.
For overdamped dynamics the respective
results are $P(x,t)=\widetilde{P}(x-v_0 t,t)$ and in Fourier-Laplace
space
\begin{equation}
\label{newtrans2}
P(k,\lambda)=\widetilde{P}(k,\lambda - i v_0 k)\:.
\end{equation}
Their evolution equations can also be shown to transform via a GT on their independent variables (Appendix~\ref{table}).
  
So far we have shown that a stochastic coarse-grained description
inherently violates GI. Nevertheless, \eqref{eq:Frule})
characterizes the stochastic dynamics in all different Galilean frames
uniquely as follows: (i) Stochastic equations of motion transform via a GT on their position and velocity processes only;
consequently, (ii) Fokker-Planck (FP) and Klein-Kramers equations also transform via a
GT on their independent variables, and (iii) PDFs transform as in
Eqs.~(\ref{newtrans1}, \ref{newtrans2}). The validity of the properties (i)--(iii) is non-trivial and needs in principle to be shown for any specific stochastic model at hand following a coarse-graining procedure. These three Galilean
transformation rules for coarse-grained stochastic dynamics and its
statistical counterparts yield what we call \textit{weak GI}: apart
from a shift of $v_0$ or $v_0t$ for velocity and position variables, respectively, the corresponding PDFs in $\widetilde{\mathcal{S}}$ remain
unchanged compared to the ones in $\mathcal{S}$. It is important to distinguish these weak GI rules from conventional microscopic GI. In systems satisfying the latter, the equations of motion are strictly identical in all inertial frames, while their stochastic coarse-grained equivalents are different. 
  
Clearly, all processes described by the generalized LE~(\ref{GLE})
satisfy (i)--(iii), which includes normal diffusive processes. In this
case the FP equation in $\widetilde{\mathcal{S}}$ is the well-known
advection-diffusion equation. 
\eqref{GLE}) also models anomalous diffusion if
one uses for $\Omega$ a power law kernel in time
\cite{kupferman2004fractional}, which highlights that these properties
are preserved in the anomalous regime. However, in modelling anomalous
diffusion a large variety of processes are used for which a similarly
rigorous coarse-graining procedure is not available
\cite{metzler2000random,klages2008anomalous,zaburdaev2015levy,KlSo11}.
While the accurate determination of an underlying anomalous stochastic
process ultimately relies on the comparison of statistical quantities
beyond the MSD with experimental data \cite{HoFr13}, we propose that
weak GI can serve as an important criterion to assess the physical
consistency of stochastic models from a purely theoretical
\textit{first principles} perspective.

In fact, we verified the validity of our conjecture for several other stochastic models generating both sub- and superdiffusion, that are commonly used in the literature, such as Fractional and Scaled Brownian motion \cite{mandelbrot1968fractional,lim2002self,hofling2013anomalous}, 
the Fractional LE \cite{Lutz2001,lim2002self,hofling2013anomalous}, 
L\'evy flights \cite{hughes1981random,fogedby1992fluctuations,chechkin2006fundamentals}, L\'evy walks \cite{shlesinger1987levy,sokolov2003towards,zaburdaev2015levy,fedotov2016single}, and the CTRW \cite{montroll1965random,fogedby1994langevin,metzler2000random}.
An overview is presented in Appendix Table~A1, where for simplicity we only demonstrate the validity of property (ii) (details of the calculations are discussed in Appendix~\ref{table}).
Remarkably, apart from the CTRW, all
representations exhibit weak GI, i.e., applying a GT to the given
Langevin or FP description yields solutions in agreement with
Eqs.~(\ref{newtrans1}, \ref{newtrans2}). For Fractional Brownian
motion, Scaled Brownian motion, and the Fractional LE (as a special
case of the generalized LE), this result can be proven based on the
Gaussian nature of the process. 
For L\'evy flights it is a direct
consequence of the L\'evy-Khintchine representation of L\'evy
processes \cite{cont1975financial}. In these examples, the Langevin dynamics can be expressed
in terms of an additive noise process and thus the transformation into
frame $\widetilde{\mathcal{S}}$ by GT is unproblematic leading to an
advective term $v_0\partial/\partial x$ as for normal diffusion. Even
though such a simple structure does not apply to L\'evy walks,
surprisingly the same consistency is satisfied, as can be checked by
imposing a GT onto the respective FP equation \cite{fedotov2016single}
and verifying that the solutions in each frame are related by
\eqref{newtrans2}. The FP equation in $\widetilde{\mathcal{S}}$
describes a L\'evy walk with asymmetric velocity jumps switching
between $-v_0+u$ and $-v_0-u$, where $\pm u$ is the velocity in
$\mathcal{S}$, which clearly is physically correct.

We now clarify the situation for the CTRW, a model that has huge
applications across all branches of the sciences
\cite{metzler2000random,mekl04,klages2008anomalous,KlSo11}. In
the overdamped regime the PDF $P$ of a CTRW in the frame $\mathcal{S}$
is the solution of the diffusion equation
\cite{magdziarz2009langevin,cairoli2015anomalous}
\begin{equation}
\label{fracdiff}
\frac{\partial}{\partial t}P(x,t)=\mathcal{L} \mathbb{D}_t P(x,t),\qquad \mathcal{L}=\sigma\frac{\partial^2}{\partial x^2},
\end{equation}
where $\sigma$ is a generalized diffusion constant and
$\mathbb{D}_t$ is a non-local time operator defined as $\mathbb{D}_t
P(x,t)=\derpar{}{t}\int_0^t{\rm
  d}t^{\prime}K(t-t^{\prime})P(x,t^{\prime})$, which generalizes the
Riemann-Liouville fractional differential operator to arbitrary
waiting time distributions. The kernel $K$ is related to the so-called
Laplace exponent $\Phi$ of the waiting time distribution by
$K(\lambda)=\Phi(\lambda)^{-1}$ \cite{magdziarz2009langevin,cairoli2015anomalous}.
Therefore, its Fourier-Laplace representation is $\mathbb{D}_t
P(x,t)\to\lambda P(k,\lambda)/\Phi(\lambda)$.  In the CTRW
framework a constant drift can be incorporated by complementing the
diffusion operator with $v_0\partial/\partial x$, which would suggest
that the FP equation in $\widetilde{\mathcal{S}}$ is given by
$\frac{\partial}{\partial
  t}\widetilde{P}=\left[v_0\frac{\partial}{\partial
    x}+\mathcal{L}\right] \mathbb{D}_t \widetilde{P}$.  
Alternatively, another time non-local FP equation was previously derived, in particular for
$\Phi(\lambda)=\lambda^\alpha$ ($0<\alpha<1$)
corresponding to L{\'e}vy stable distributed waiting times, 
by employing the transformation rule \eqref{newtrans2} and performing a Taylor expansion in the Fourier variable up to the lowest approximation order 
\cite{metzler1998anomalous,metzler1999transport,metzler2000random}. 
This procedure leads to the equation:  
$\frac{\partial}{\partial t}\widetilde{P}=v_0\frac{\partial}{\partial
  x}\widetilde{P}+\mathcal{L} \mathbb{D}_t \widetilde{P}$.

\begin{figure*}[!ht]
\centering
\includegraphics[width=160mm]{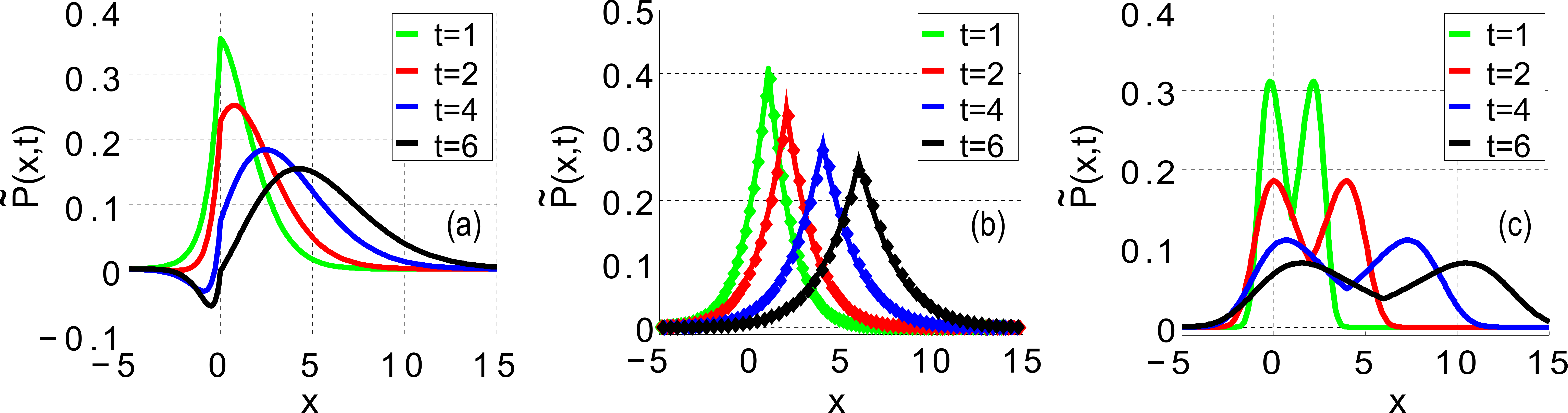}
\caption{Position distribution in the comoving frame
  $\widetilde{\mathcal{S}}$.  (a) Propagator of the Fokker-Planck equation
  $\frac{\partial}{\partial
    t}\widetilde{P}=v_0\frac{\partial}{\partial
    x}\widetilde{P}+\mathcal{L} \mathbb{D}_t \widetilde{P}$ suggested
  in \cite{metzler1998anomalous,metzler1999transport,metzler2000random} instead of
  \eqref{fracad}. The explicit expression for the propagator is given in {Appendix~\ref{propagator}}, Eq.~(S47). 
  This function not only violates weak GI, but also
  exhibits non-physical negative values. Here,
  $\Phi(\lambda)=\lambda^{\alpha}$ with $\alpha=0.5$,
  $v_0=-1$, $\sigma=1$, $x_0=0$.  (b) PDF solution of
  \eqref{fracad} {(Appendix~\ref{xibar}, Eq.~(S68))} showing weak GI. Parameters are the same as
  for (a). We find perfect agreement with Monte-Carlo
  simulations of the Langevin equation $\dot{\widetilde{Y}}(t)=-v_0 +
  \overline{\xi}(t)$ (colored markers).  (c) PDF solution of
  \eqref{fracad} {(Appendix~\ref{xibar}, Eq.~(S68))} with $K(t)=t^{\alpha-1}/\Gamma(\alpha)$ and
    $\alpha$ analytically continued to yield superdiffusion at
    $\alpha=1.5$. Other parameters as for (b). Again, weak GI
  is observed.}
  \label{mkpdf}
\end{figure*}

However, both equations are not correct representations of microscopic
dynamics in view of the rules (i)--(iii) yielding weak GI.  In fact,
the former does not satisfy the general rule \eqref{newtrans2} as
becomes clear by solving it in Fourier-Laplace space. The same is true
for the latter, whose solutions are even unphysical, as they do not
satisfy the requirement of positivity of a PDF (Fig.~\ref{mkpdf}a and Appendix~\ref{propagator}).  Therefore, a
simple transformation of the fractional diffusion equation obtained by
arbitrarily adding an advective term $v_0\partial/\partial x$ as for
the Gaussian models and L\'evy flights (see Table~S1) is not
correct. Likewise, implementing GTs directly on the Langevin
description of CTRWs in terms of subordination
\cite{fogedby1994langevin,cairoli2015anomalous,cairoli2017timedep} is
problematic {(see below)}.

Instead, the correct transformation of \eqref{fracdiff}) into the frame
$\widetilde{S}$ can be derived straightforwardly in Fourier-Laplace space.   
Without loss of generality, we assume $P(x,0)=\delta(x)$. 
Thus, its transform is   
$\lambda P(k,\lambda)-1=-\sigma k^2[\lambda/\Phi(\lambda)]P(k,\lambda)$.
Employing property (iii), the GT is then implemented by the variable transformation 
$\lambda \to \lambda + i k v_0$ and the transformation rule \eqref{newtrans2}) relating $P, \widetilde{P}$. 
This immediately leads to a FP equation including retardation
effects
\begin{equation}
\label{fracad}
\frac{\partial}{\partial t}\widetilde{P}(x,t)=v_0\frac{\partial}{\partial x}\widetilde{P}(x,t)+\mathcal{L} \mathcal{D}_t^{(v_0)} \widetilde{P}(x,t)\:,
\end{equation}
where the operator  
$\mathcal{D}^{(v_0)}_t$ is the fractional substantial derivative \cite{sokolov2003towards,Friedrich2006Anomalous,cairoli2015anomalous}
\begin{equation}
\mathcal{D}^{(v_0)}_t \widetilde{P}(x,t)=\left[ \frac{\partial}{\partial t}-v_0\frac{\partial}{\partial x}\right]\int_0^t{\rm d}t^{\prime} K(t-t^{\prime})\widetilde{P}(x+v_0(t-t^{\prime}),t^{\prime}) 
\label{eq:subDeriv}
\end{equation}
which has Fourier-Laplace representation
$\mathcal{D}^{(v_0)}_t
\widetilde{P}(x,t) \to (\lambda +i v_0
k)\widetilde{P}(k,\lambda)/\Phi(\lambda + i v_0 k)$.
Setting $v_0\!=\!0$ recovers \eqref{fracdiff}. 

To further support our result, we also derive \eqref{fracad} directly in $(x,t)$-space. This requires a careful analysis due to the non-local character of the operator $\mathbb{D}_t$. 
On the one hand, the lhs of
\eqref{fracdiff} and the time derivative in front of $\mathbb{D}_t$
transform with the substitution $\partial/\partial t
\to \partial/\partial t -v_0 \partial/\partial x$ (chain rule
applied to Eqs.~(\ref{eq:Gtransf})).  On the other hand, recalling the
explicit definition of a PDF in terms of probability (denoted as
$\mathds{P}$) of events (denoted as $\{\cdot\}$), the integrand PDF is
defined as $P(x,t^{\prime})=\mathds{P}(\{x \leq Y(t^{\prime}) \leq
x+\diff{x}\})$, {where $Y(t)$ denotes the position of the CTRW}.  
According to property (i), $Y(t^{\prime})$ becomes $\widetilde{Y}(t^{\prime})+v_0 t^{\prime}$ in the comoving frame $\widetilde{\mathcal{S}}$, while the measured position $x$ transforms at the later time $t$ in agreement with the lhs of the equation, i.e., $x \!\to\! x+v_0
t$. Therefore $P(x,t^{\prime})=\mathds{P}(\{x+v_0(t-t^{\prime})
\leq \widetilde{Y}(t^{\prime}) \leq
x+v_0(t-t^{\prime})+\diff{x}\})=\widetilde{P}(x+v_0(t-t^{\prime}),t^{\prime})$. 
Note that $\diff{x}$ is invariant because the shift cancels out. 
Combining these arguments yields \eqref{fracad}. The fractional
substantial derivative in \eqref{eq:subDeriv} highlights the
existence of a space-time coupling, which is absent in the frame
$\mathcal{S}$ but is naturally required: let $y$ be the position of
the CTRW in $\mathcal{S}$ after its last jump occurred at time $t$,
and $\tau$, $\Delta y$ respectively the waiting time to the next jump
and its length. In $\mathcal{S}$ its position at time $t+\tau$ is then
$y+\Delta y$. In $\widetilde{\mathcal{S}}$ this is 
$y-v_0\tau+\Delta y$ (\eqref{eq:Gtransf}, left). Thus, the
final position in $\widetilde{\mathcal{S}}$ depends on both the jump
amplitude $\Delta y$ and the waiting time $\tau$. 
Interestingly, a similar coupling is constitutive of the L{\'e}vy walk model \cite{zaburdaev2015levy}, which explains why it satisfies weak GI.

What is now the corresponding Langevin dynamics of the anomalous
diffusive process described by \eqref{fracad}? The key is to
describe the CTRW directly in physical time rather than in the widely
used subordination picture
\cite{fogedby1994langevin,cairoli2015anomalous,cairoli2017timedep}. 
In the physical time representation a CTRW in $\mathcal{S}$ is given as
$\dot{Y}(t)=\overline{\xi}(t)$, where $\overline{\xi}$ is the derivative of a subordinated Brownian motion \cite{cairoli2015langevin}.  
This is equivalently written as the formal definition 
$\overline{\xi}(t)=\int_0^{\infty} \xi(s)\delta(t-T(s))$, 
where $\xi$ is a white Gaussian noise with $\langle \xi(t)\rangle=0$ and $\langle \xi(t_1)\xi(t_2)\rangle=2\sigma\delta(t_1-t_2)$, and $T$ is a strictly increasing L{\'e}vy process. 
Using this representation, we can calculate the characteristic functional $G$ of $\overline{\xi}$ for a general test function $u$ (Appendix~\ref{xibar}) 
\begin{align}
G[u(r)]&=\Average{\exp{\left[ -\sigma \int_0^{\infty} [u(T(s))]^2 \diff{s}\right]}} ,\label{charF}
\end{align}  
where the brackets denote an average over the realizations of the process $T$.  
A GT can now be performed without problems
leading to $\dot{\widetilde{Y}}(t)=-v_0 + \overline{\xi}(t)$.
Remarkably, employing functional techniques \cite{caceres1997generalized} together with the result in \eqref{charF}, we can show that the FP equation for this process is precisely given by
\eqref{fracad}, thus completing the picture (Appendix~\ref{xibar}).
The Langevin description in physical
time highlights that to correctly implement the change of frame, the
constant advective force exerted on the underlying random walk in the
frame $\widetilde{S}$ needs to act at each time step, i.e., also
during the trapping times. This simple physical scenario underlies the
complicated space-time coupling manifest in the retardation of
\eqref{fracad}. Its modelling in terms of subordination thus
inevitably couples the equations for the position and elapsed time
processes, which makes any analytical treatment challenging
(an example is discussed in Appendix~\ref{subordination}, where we derive \eqref{fracad} for the process $\widetilde{Y}$ using its representation in terms of coupled subordinated equations). 
Further using the characteristic functional of the noise $\overline{\xi}$ in \eqref{charF} one can derive its analytical solution 
\begin{equation}
\widetilde{P}(k,\lambda)=\frac{1}{\lambda + i k v_0} \left[ 1- \frac{\sigma k^2}{\Phi(\lambda + i k v_0)+\sigma k^2} \right],
\label{eq:PDFcostF}
\end{equation}
whose inverse Fourier-Laplace transform is plotted in Fig.~\ref{mkpdf}b for the particular case of $T$ being a L{\'e}vy stable process of order $\alpha$ (Appendix~\ref{xibar}, Eq.~(S68)). 
We observe the typical distribution of a force-free CTRW \cite{metzler2000random}
time-shifted with velocity $v_0$, in perfect agreement with
numerical simulations of $\widetilde{Y}$.

Moreover, we find that $\widetilde{Y}$ can also generate a superdiffusive MSD thus providing a unified model for both sub- and
superdiffusion. This surprising fact relies on the Langevin
description in physical time and the equivalent characterization of
$\overline{\xi}$ by means of its multipoint correlation functions
\cite{cairoli2015langevin}.  In particular, its FP equation is still
\eqref{fracad}, which can be derived by a generalization of
  Novikov's theorem via functional methods
  \cite{novikov1965functionals,hanggi1978correlation} (Appendix~\ref{superdiff}), and the resulting PDF satisfies weak
GI. In Fig.~\ref{mkpdf}c we plot its propagator for $K(t)=t^{\alpha-1}/\Gamma(\alpha)$, now for $1<\alpha<2$ (Appendix~\ref{xibar}, Eq.~(S68), analytically continued in $\alpha$). For $v_0=0$, this PDF was discussed in \cite{metzler2000accelerating}.

In summary, using a Galilean invariant version of the
  paradigmatic Kac-Zwanzig model, we have derived the weak GI
  properties (i)--(iii) that need to be satisfied in order to
  consistently describe the same stochastic system in different
  inertial frames. While these properties hold for normal diffusion
  based on our analytical derivation, by employing these rules
  consistent anomalous diffusive models can be constructed for both
  sub- and superdiffusion, even though a precise coarse-graining
  procedure is missing for them. We demonstrated this by providing
  the missing representation for the important class of CTRW models,
  which shows that the correct form is not at all suggested from the
  representation in the rest frame. Moreover, the Langevin
  representation (i) discloses that in a comoving frame the heat bath
  leads generally to an additive flow field on the tracer particle
  irrespective of the details of the underlying coupling.
  Consequently, the definitions of work, heat and entropy production
  used within the recent theory of stochastic thermodynamics
  \cite{Seifert:2012aa} have to be modified to account for the
  contribution of the external flow \cite{Speck:2008aa} highlighting
  fundamental similarities between normal and anomalous diffusive
  systems, even though the stochastic thermodynamics of the latter is
  so far not well understood \cite{Chechkin:2012aa}. Along these
  lines, connections between GI and the validity of
  fluctuation-dissipation relations on the one hand, and the
  celebrated fluctuation relations generalizing the second law of
  thermodynamics \cite{Seifert:2012aa} on the other, have been
  suggested \cite{Chechkin:2012aa,Dieterich:2015aa} and need to be
  investigated further. But our most important statement is that
  ignoring our weak GI rules can easily lead to unphysical models, as
  exemplified by the CTRW with an \textit{ad hoc} advective term
  (Fig.~\ref{mkpdf}a). The consequences of our results are thus
  far-reaching. Weak GI is expected to constrain all mesoscopic
  diffusive models whose microscopic representation is expected to
  satisfy conventional GI. As such, it provides an important selection
  principle on stochastic models preceding comparison with data, which
  can guide modelling approaches throughout the physical, chemical,
  and biological sciences.


\acknowledgments{A.C. gratefully acknowledges funding under the Postgraduate
  Research Fund (QMPGRF) granted by Queen Mary University of London
  and under the Science Research Fellowship granted by the Royal
  Commission for the Exhibition of 1851. R.K. thanks the Office of Naval Research Global for financial support. He also acknowledges funding from the London Mathematical Laboratory, where he is an External Fellow. A.B. gratefully acknowledges funding under EPSRC grant EP/L020955/1. We thank A.V.~Chechkin for fruitful discussions and for technical support of the derivation presented in Appendix~\ref{propagator}.
}



\onecolumngrid
\appendix

The Appendices are organized as follows. 
In appendix~\ref{rule}, we derive the transformation rule between different inertial frames $\mathcal{S}$ and $\widetilde{\mathcal{S}}$, moving at relative velocity $v_0$, of position and velocity processes satisfying the generalized Langevin equation (LE) (Eq.~\eqref{GLE}. 
This is obtained by only employing the transformation rule of their stochastic equations of motion, that we derive analytically from the Kac-Zwanzig model (main text). This calculation thus provides a derivation of the transformation rule for their joint statistics Eq.~\eqref{newtrans1}. 
Appendix~\ref{table} contains detailed derivations of the Fokker-Planck (FP) type equations in both frames $\mathcal{S}$ and $\widetilde{\mathcal{S}}$, that are shown in Table~A1, for several stochastic processes generating both normal and anomalous diffusion. 
In particular, we discuss overdamped Gaussian processes, the generalized LE, and the L{\'e}vy walk. This discussion highlights that weak GI is indeed satisfied by all such processes.   
In appendix~\ref{propagator}, we derive analytically the propagator of the incorrect FP equation of a continuous-time random walk (CTRW) in the comoving frame $\widetilde{\mathcal{S}}$, originally proposed in refs.~\cite{metzler1998anomalous,metzler1999transport,metzler2000random}, which is numerically plotted in Fig.~\ref{mkpdf}a. 
In appendix~\ref{xibar}, we derive the characteristic functional of the noise $\overline{\xi}$, which is defined as the time derivative of a subordinated Brownian motion. We then use this result to verify that the FP equation of a process $X$, whose dynamics is described by the LE $\dot{X}=-v_0+\overline{\xi}$, is the non local advection-diffusion Eq.~\eqref{fracad}. 
In appendix~\ref{subordination}, we provide an alternative derivation that employs the formulation of a CTRW in terms of subordinated processes. This discussion elucidates the effect of the spatio-temporal coupling imposed by weak GI on the subordinated LEs.  
In appendix~\ref{superdiff}, we show that $\overline{\xi}$ can be used to describe more general processes, including superdiffusive ones, that do not possess a formulation in terms of subordination. We then give a proof that their FP equation is still Eq.~\eqref{fracad}.    
Appendix~\ref{specialF} contains a technical note about the Fox H-function and the three parameter Mittag Leffler function, whose properties are used throughout the main text and SI. Below we denote with X,V position and velocity
processes for general dynamics, except for the CTRW whose position is called Y.

\begin{table}[!htb]
\centering
\begin{tabular*}{\hsize}{@{}lll}
Stochastic model $\qquad$ & Fokker-Planck/Klein-Kramers eq. in $\mathcal{S}$ & Fokker-Planck/Klein-Kramers eq. in $\widetilde{\mathcal{S}}$ \\
\hline

\vspace{-0.8mm}&\vspace{-0.8mm}&\vspace{-0.8mm}\\ 

Normal diffusion (overdamped) & $\left[ \frac{\partial}{\partial t} -\mathcal{L}\right]P=0\qquad$ & $\left[\frac{\partial}{\partial t}-v_0\frac{\partial}{\partial x}-\mathcal{L}\right]\widetilde{P}=0$    \\

\vspace{-0.8mm}&\vspace{-0.8mm}&\vspace{-0.8mm}\\ 

Normal diffusion (underdamped) 
\footnote{$\gamma\!>\!0$ is the friction coefficient.}

& $\left[ \frac{\partial}{\partial t}+\derpar{}{x}v-\derpar{}{v}\gamma v - \gamma\sigma\dersecpar{}{v} \right]P=0 $ & $\left[ \frac{\partial}{\partial t}+\derpar{}{x}v-\derpar{}{v}\gamma (v+v_0) - \gamma\sigma\dersecpar{}{v} \right]\widetilde{P}=0$    \\

\vspace{-0.8mm}&\vspace{-0.8mm}&\vspace{-0.8mm}\\ 

Fractional/Scaled Brownian motion
\footnote{$0\!<\!\beta\!<\!2$ is the exponent of the characteristic power-law dependence of the noise correlations.}

 & $\left[\frac{\partial}{\partial t}-\beta t^{\beta -1}\mathcal{L}\right]P=0$ & $\left[\frac{\partial}{\partial t}-v_0 \derpar{}{x}-\beta t^{\beta -1}\mathcal{L}\right]\widetilde{P}=0$ \\

\vspace{-0.8mm}&\vspace{-0.8mm}&\vspace{-0.8mm}\\ 

Generalized Langevin equation 
\footnote{$\Gamma$, $D_{xv}$ are time dependent friction and diffusion coefficients, respectively, given in Eq.~\eqref{GLEcoeff}.}

& $\left[ \derpar{}{t} + \derpar{}{x} v - \derpar{}{v} \Gamma(t) v \right]P$ & $\left[ \derpar{}{t} + \derpar{}{x} v - \derpar{}{v}\Gamma(t)(v+v_0)\right]\widetilde{P}$ \\
\vspace{-0.85mm}&\vspace{-0.85mm}&\vspace{-0.85mm}\\ 
& $\qquad =\left[ \dersecpar{}{v}\sigma\Gamma(t) + \frac{\partial^2}{\partial x \partial v}D_{xv}(t)\right]P$ & $\qquad =\left[ \dersecpar{}{v}\sigma\Gamma(t) + \frac{\partial^2}{\partial x \partial v}D_{xv}(t)\right]\widetilde{P}$ \\

\vspace{-0.8mm}&\vspace{-0.8mm}&\vspace{-0.8mm}\\ 

L\'evy flight
\footnote{$\nabla^\beta$ ($0\!<\!\beta\!<\!2$) denotes the fractional Laplacian, defined in Fourier space as $\nabla^{\beta}\!\to\! -|k|^{\beta}$.}

& $\left[\frac{\partial}{\partial t}-\nabla^\beta\right] P=0$ & $\left[\frac{\partial}{\partial t}-v_0\frac{\partial}{\partial x}-\nabla^\beta\right]\widetilde{P}=0$ \\


\vspace{-0.8mm}&\vspace{-0.8mm}&\vspace{-0.8mm}\\ 
  
L\'evy walk
\footnote{$u$ is the absolute value of the velocity in the frame $\mathcal{S}$, while in $\widetilde{\mathcal{S}}$ the forward/backward velocities are $u_{\pm}=-v_0\pm u$. The operator $\mathcal{D}_t^{(v_1,v_2)}$ has the representation $\mathcal{D}^{(v_1,v_2)}_t P(x,t)\to(\lambda - i k v_1)K(\lambda - i k v_2)P(k,\lambda)$ (see Eq.~\eqref{LWfracop}). For $v_1=v_2=-v_0$, $\mathcal{D}^{(v_1,v_2)}_t$ recovers the fractional substantial derivative Eq.~\eqref{eq:subDeriv}.}

& $\left[\left(\frac{\partial}{\partial t}+u \derpar{}{x}\right)\left(\frac{\partial}{\partial t}-u \derpar{}{x}\right)\right] P_u$ & $\left[\left(\frac{\partial}{\partial t}+u_+\derpar{}{x}\right)\left(\frac{\partial}{\partial t}+u_-\derpar{}{x}\right)\right]\widetilde{P}_u$\\
\vspace{-0.85mm}&\vspace{-0.85mm}&\vspace{-0.85mm}\\ 
& $\qquad =-\left[\frac{1}{2}\mathcal{D}_t^{(-u,u)} +\frac{1}{2}\mathcal{D}_t^{(u,-u)}\right] P_u$ & $\qquad =-\left[\frac{1}{2}\mathcal{D}_t^{(u_-,u_+)} +\frac{1}{2}\mathcal{D}_t^{(u_+,u_-)}\right] \widetilde{P}_u$\\

\vspace{-0.8mm}&\vspace{-0.8mm}&\vspace{-0.8mm}\\ 

Continuous time random walk & $\left[\frac{\partial}{\partial t}-\mathcal{L} \mathbb{D}_t \right] P=0$  & \multicolumn{1}{c}{?} \\

\vspace{-0.8mm}&\vspace{-0.8mm}&\vspace{-0.8mm}\\
\hline
\end{tabular*}
\caption{Overview of generic stochastic models for normal and anomalous diffusion. For simplicity, we show their representations in terms of generalized Fokker-Planck or Klein-Kramers equations and neglect the explicit dependencies of the distributions $P$, $\widetilde{P}$ on the sample variables.
For all models, except the Continuous time random walk, property $\textit ii$ holds, i.e., their evolution equations in different inertial frames are related by a Galilean transformation of their independent variables.
We define the diffusion operator $\mathcal{L}=\sigma\frac{\partial^2}{\partial x^2}$. }\label{tableS1}




%
\end{table}

\section{Solution of the generalized Langevin equations in $\mathcal{S}$ and $\widetilde{\mathcal{S}}$  \label{rule}}
  
Let us consider the generalised LE in the laboratory frame $\mathcal{S}$:
\begin{align}
\dot{X}(t)&=V(t), &
M \dot{V}(t)&=-\int_0^t \Omega(t-t^{\prime})V(t^{\prime})\diff{t^{\prime}} + \xi(t) . 
\label{gleS}
\end{align}  
The tracer trajectory $(X(t),V(t))$, with initial condition $(X_0,V_0)$ at time $t=0$, can be obtained exactly by Laplace transforming Eq.~\eqref{gleS}. For the position, this yields 
$\lambda X(\lambda)-X_0=V(\lambda)$, 
while for the velocity 
\begin{equation}
V(\lambda)=\frac{M V_0}{M \lambda +\Omega(\lambda)}+\frac{\xi(\lambda)}{M \lambda +\Omega(\lambda)} . 
\end{equation} 
Transforming back these equations in time space, we obtain 
$X(t)=X_0+\int_0^t V(t^{\prime}) \diff{t^{\prime}}$ 
and
\begin{equation}
V(t)=M V_0 w(t)+\int_0^t w(t-t^{\prime})\xi(t^{\prime})\diff{t^{\prime}}, 
\label{exactVgle}
\end{equation} 
where the function $w$ is defined in Laplace transform by 
\begin{equation}
w(\lambda)=[M\lambda +\Omega(\lambda)]^{-1}.
\label{GLEtkernel}
\end{equation} 
We then consider the corresponding dynamics in the comoving frame $\widetilde{\mathcal{S}}$. These are described by  
\begin{align}
\dot{\widetilde{X}}(t)&=\widetilde{V}(t), &
M \dot{\widetilde{V}}(t)&=-\int_0^t \Omega(t-t^{\prime})[\widetilde{V}(t^{\prime})+v_0]\diff{t^{\prime}}+\xi(t) . 
\label{gleStilde}
\end{align}
As before, we can derive the exact trajectory $(\widetilde{X}(t),\widetilde{V}(t))$ by taking the Laplace transform of Eqs.~\eqref{gleStilde}. This yields for the position  
$\lambda \widetilde{X}(\lambda)-\widetilde{X}_0=\widetilde{V}(\lambda)$ 
and for the velocity 
\begin{equation}
\widetilde{V}(\lambda)=\frac{M \widetilde{V}_0}{M \lambda +\Omega(\lambda)}-\frac{v_0\Omega(\lambda)}{\lambda[M \lambda +\Omega(\lambda)]}+\frac{\xi(\lambda)}{M \lambda +\Omega(\lambda)} ,  
\label{exactYtilde}
\end{equation} 
where $\widetilde{X}_0,\widetilde{V}_0$ are the initial condition in the transformed frame.   
Employing the relations: $\widetilde{V}_0=V_0-v_0$ and $\widetilde{X}_0=X_0$, that result from the Galilean transformation (GT) Eq.~\eqref{eq:Gtransf}, 
we find 
\begin{align}
\widetilde{V}(\lambda)&=\frac{M (V_0-v_0)}{M \lambda +\Omega(\lambda)}-v_0\frac{\Omega(\lambda)}{\lambda[M \lambda +\Omega(\lambda)]}+\frac{\xi(\lambda)}{M \lambda +\Omega(\lambda)} \notag\\
&=\frac{M V_0}{M \lambda +\Omega(\lambda)}+\frac{\xi(\lambda)}{M \lambda +\Omega(\lambda)}-\frac{v_0}{\lambda} \notag\\
&
=V(\lambda)-\frac{v_0}{\lambda} .  
\end{align}
Substituting this equation into that of the position, we can write 
\begin{align}
\lambda \widetilde{X}(\lambda)&=X_0+V(\lambda)-\frac{v_0}{\lambda} 
=\lambda X(\lambda)-\frac{v_0}{\lambda} . 
\end{align}
Taking their inverse Laplace transforms yields: $\widetilde{V}(t)=V(t)-v_0$ and $\widetilde{X}(t)=X(t)-v_0 t$. These transformation rules for $X,V$ directly provide Eq.~\eqref{newtrans1}.

\section{Analysis of weak Galilean invariance for several stochastic coarse-grained models \label{table}}

In this appendix, we study several different stochastic models \cite{mandelbrot1968fractional,lim2002self,hofling2013anomalous,Lutz2001,shlesinger1987levy,sokolov2003towards,zaburdaev2015levy,fedotov2016single,montroll1965random,fogedby1994langevin,metzler2000random}, that are widely used in the literature to model both normal and anomalous diffusion, in terms of weak Galilean invariance (GI). An overview is given in Table~\ref{tableS1}.

\subsection{Overdamped Gaussian processes: fractional and scaled Brownian motion}
\label{sifsbm}

General overdamped Gaussian processes are described in the laboratory frame $\mathcal{S}$ by the LE
\begin{equation}
\dot{X}(t)=\xi(t), 
\label{eq:BalescuL}
\end{equation}
where $\xi(t)$ is a Gaussian coloured noise with $\Average{\xi(t)}=0$ and two-point correlation function
$\Average{\xi(t)\xi(t^{\prime})}=C(t,t^{\prime})$. 
The time evolution of its position distribution $P(x,t)=\Average{\delta(x-X(t))}$ is given by
\begin{equation}
\derpar{}{t}P(x,t)=-\derpar{}{x}\left\langle \xi(t) \delta(x-X(t)) \right\rangle .
\label{eq:GaussCorrFP2}
\end{equation}  
To get a closed equation for $P$, one needs to compute the averaged quantity in its right-hand side (rhs).
For Gaussian noise, one employs Novikov's theorem \cite{novikov1965functionals,hanggi1978correlation}, that yields 
\begin{align}
\left\langle \xi(t) \delta(x-X(t)) \right\rangle &\!= \!\int_0^t C(t,t^{\prime}) \left\langle \frac{\delta[\delta(x-X(t))]}{\delta \xi\!\left(t^{\prime}\right)}  \right\rangle \diff{t^{\prime}} \notag\\
&=-\derpar{}{x}\int_0^t C(t,t^{\prime})\left\langle \delta(x-X(t)) \frac{\delta X(t)}{\delta \xi(t^{\prime})} \right\rangle \diff{t^{\prime}}=-\derpar{}{x} D(t)P(x,t),
\label{eq:Novikov}
\end{align}
where $\delta X(t)/\delta \xi(t^{\prime})=\Theta\!\left(t-t^{\prime}\right)$ and 
$D(t)=\int_0^t C(t,t^{\prime})\diff{t^{\prime}}$. 
Substituting it in Eq.~\eqref{eq:GaussCorrFP2}, we obtain 
\begin{equation}
\derpar{}{t}P(x,t)=\dersecpar{}{x}D(t)P(x,t). 
\label{eq:GaussCorrFP}
\end{equation}  
The previous argument holds for both stationary noises, whose correlation function depends only on the time difference, i.e., $C(t,t^{\prime})=C(|t-t^{\prime}|)$, and non-stationary ones.    
In the former case, an important example is the fractional Brownian motion; 
in the latter case, the scaled Brownian motion \cite{mandelbrot1968fractional,lim2002self,hofling2013anomalous}.
These processes are defined by setting the two-point correlation function equal to 
$C(|t-t^{\prime}|)=\beta(\beta-1)|t-t^{\prime}|^{\beta-2}$ and 
$C(t,t^{\prime})=\beta t^{\beta-1}\delta(t-t^{\prime})$ 
respectively with $0<\beta<2$ \cite{hofling2013anomalous}, 
that yield the same diffusion coefficient       
$D(t)=\beta t^{\beta-1}$.  
Eq.~\eqref{eq:GaussCorrFP} is easily solved by the Gaussian  
$P(x,t)=e^{-\frac{x^2}{4 \Sigma(t)}}/\sqrt{4\pi\Sigma(t)}$, 
where $\Sigma(t)=\int_0^t D(t^{\prime})\diff{t^{\prime}}$. 
Applying the GT Eq.~\textbf{1}, we obtain: 
$\widetilde{P}(x,t)=e^{-\frac{(x+v_0 t)^2}{4 \Sigma(t)}}/\sqrt{4\pi\Sigma(t)}$. This is easily shown to satisfy the FP equation: 
\begin{equation}
\derpar{}{t}\widetilde{P}(x,t)=\left[ \derpar{}{x} v_0 + \dersecpar{}{x}D(t)\right] \widetilde{P}(x,t),  
\end{equation}
that corresponds to the LE  
\begin{equation}
\dot{\widetilde{X}}(t)=-v_0+\xi(t).  
\end{equation}
Therefore, the description of overdamped Gaussian processes satisfies properties $i\!-\!iii$ (main text), i.e., it exhibits weak GI. 


\subsection{Generalised Langevin equation\label{sigle}}

We write the generalised Langevin Eq.~\eqref{GLE} 
as (we set $M=1$ without loss of generality) 
\begin{align}
\dot{X}(t)&=V(t), & 
\dot{V}(t)&=-\int_0^t \Omega(t-t^{\prime})V(t^{\prime})\diff{t^{\prime}}+\xi(t), 
\label{fracLE}
\end{align}
where $\Omega$ is a prescribed drag coefficient and the coloured Gaussian noise $\xi$ has the two point correlation function 
\begin{equation}
\langle \xi(t)\xi(t^{\prime})\rangle=C(|t-t^{\prime}|)=\sigma \Omega(|t-t^{\prime}|) , 
\end{equation} 
with $\sigma=k_B T$ ($T$ is the temperature of the bath at equilibrium). 
Thus, it satisfies the fluctuation-dissipation relation \cite{kubo1966fluctuation}.
Relevant examples are (a) underdamped normal diffusion, for which $\Omega(t)=\gamma\delta(t)$ ($\gamma>0$), and (b) fractional LE \cite{Lutz2001,lim2002self,hofling2013anomalous}, for which $\Omega(t)=\gamma_{\alpha}t^{-\alpha}/\Gamma(1-\alpha), \,\, 0\!<\!\alpha\!<\!1, \,\, \gamma_{\alpha}>0$.
We call $X_0=X(0), V_0=V(0)$ the initial conditions. Eq.~\eqref{fracLE} has been widely discussed in the main text in terms of weak GI. In particular, the validity of the properties $i, iii$ has been discussed. 
Here, we show that also property $ii$ holds.
First, we derive the Klein-Kramers equation for its joint position-velocity probability density function (PDF) in the laboratory frame $\mathcal{S}$ $P(x,v,t)=\langle \delta(x-X(t))\delta(v-V(t))\rangle$.
Due to the Gaussian nature of $\xi$, and using the exact solution of the dynamics Eq.~\eqref{exactVgle}, the joint characteristic function is \cite{adelman1976fokker,wang1999nonequilibrium}
\begin{equation}
P(k,p,t)=\exp{\left\{i \langle X(t) \rangle k + i \langle V(t) \rangle p -\frac{1}{2}[\sigma^2_{xx}(t)k^2 + 2\sigma^2_{xv}(t)k p + \sigma^2_{vv}(t) p^2] \right\}}, 
\end{equation}
where 
$\langle X(t) \rangle=V_0 \overline{w}(t)+X_0$,
$\langle V(t) \rangle=V_0 w(t)$ and 
we defined the auxiliary function 
$\overline{w}(t)=\int_0^t w(t^{\prime})\diff{t^{\prime}}$
and  
\begin{align}
\sigma^2_{xx}(t)&=\sigma\left[2\int_0^t \overline{w}(t^{\prime})\diff{t^{\prime}}-\overline{w}^2(t)\right] , &
\sigma^2_{vv}(t)&=\sigma[1-w^2(t)] , &
\sigma^2_{xv}(t)&=\sigma\overline{w}(t)[1-w(t)].
\end{align}
We take the following partial derivatives in $t,p$ (to ease notation we drop any explicit dependence of $P$ on its variables): 
\begin{subequations}
\begin{align}
\frac{1}{P}\derpar{}{t}P&=i V_0[w(t)k+\dot{w}(t)p]-\frac{1}{2}\left[2 \sigma^2_{xv}(t)k^2 + 2\dertot{}{t}\sigma^2_{xv}(t) k p+\dertot{}{t}\sigma^2_{vv}(t)p^2\right], \label{dPdt}\\
\frac{1}{P}\derpar{}{p}P&=-\sigma^2_{vv}(t)p-\sigma^2_{xv}(t)k + i V_0 w(t) , \label{dPdp}  
\end{align}
\end{subequations}
where we further used the relation $\dertot{}{t}\sigma_{xx}^2=2\sigma_{xv}^2$. 
Eliminating $V_0$, we derive the following equation: 
\begin{align}
\frac{1}{P}\derpar{}{t} P&=[k-\Gamma(t)p] \frac{1}{P}\derpar{}{p}P-D_{vv}(t)p^2-D_{xv}(t)k p, 
\end{align}
where the drag and diffusion coefficients $\Gamma, D_{vv}, D_{xv}$ are defined as 
\begin{align}
\Gamma(t)&=-\frac{\dot{w}(t)}{w(t)} , &
D_{vv}(t)&=\sigma\Gamma(t) , & 
D_{xv}(t)&=\sigma[-1+w(t)+\Gamma(t)\overline{w}(t)].
\label{GLEcoeff}
\end{align}
Taking its inverse Fourier transform yields: 
\begin{align}
\derpar{}{t} P(x,v,t)&=\left[ -\derpar{}{x} v + \derpar{}{v}\Gamma(t)v + \dersecpar{}{v}\sigma\Gamma(t) + \frac{\partial^2}{\partial x \partial v}D_{xv}(t)\right]P(x,v,t) .
\label{FLEkk}
\end{align}
For example (a), we find  
$\Gamma(t)=\gamma$, $D_{xv}(t)=0$,   
thus yielding the ordinary Klein-Kramers equation:
\begin{align}
\derpar{}{t} P(x,v,t)&=-\derpar{}{x} v P(x,v,t) + \gamma \derpar{}{v}\left[ v + \sigma\derpar{}{v} \right]P(x,v,t) .   
\end{align}

Let us now consider the generalized LE in the comoving frame $\widetilde{\mathcal{S}}$, i.e., Eq.~\eqref{GLEtilde}
, which we write as  
\begin{align}
\dot{\widetilde{X}}(t)&=\widetilde{V}(t), & 
\dot{\widetilde{V}}(t)&=-\int_0^t \Omega(t-t^{\prime})[\widetilde{V}(t^{\prime})+v_0]\diff{t^{\prime}}+\xi(t).  
\label{si_GLEtilde}
\end{align}
We now apply the previous technique to compute its Klein-Kramers equation. 
Being related by the GT Eq.~\eqref{eq:Gtransf}
, only their first moment changes to 
$\langle \widetilde{X}(t) \rangle=\langle X(t) \rangle -v_0 t$,
$\langle \widetilde{V}(t) \rangle=\langle V(t) \rangle -v_0$. 
Therefore, the joint characteristic function in $\widetilde{\mathcal{S}}$ is 
\begin{equation}
\widetilde{P}(k,p,t)=\exp{\left\{i \langle X(t) \rangle k + i \langle V(t) \rangle p -i k v_0 t - i p v_0 -\frac{1}{2}[\sigma^2_{xx}(t)k^2 + 2\sigma^2_{xv}(t)k p + \sigma^2_{vv}(t) p^2] \right\}}, 
\end{equation}
such that Eqs.~\eqref{dPdt}, \eqref{dPdp} changes to 
\begin{subequations}
\begin{align}
\frac{1}{\widetilde{P}}\derpar{}{t}\widetilde{P}&=i V_0[w(t)k+\dot{w}(t)p] - i k v_0 -\frac{1}{2}[2 \sigma^2_{xv}(t)k^2 + \dot{\sigma}^2_{xv}(t) k p+\dot{\sigma}^2_{vv}(t)p^2], \label{dPdt_tilde}\\
\frac{1}{\widetilde{P}}\derpar{}{p}\widetilde{P}&=-\sigma^2_{vv}(t)p-\sigma^2_{xv}(t)k -i v_0 + i V_0 w(t). \label{dPdp_tilde}  
\end{align}
\end{subequations}
Elimination of the parameter $V_0$ yields 
\begin{align}
\frac{1}{\widetilde{P}}\derpar{}{t} \widetilde{P}&=[k-\Gamma(t)p] \frac{1}{\widetilde{P}}\derpar{}{p}\widetilde{P} - i p \Gamma(t) v_0 - D_{vv}(t)p^2 - D_{xv}(t)k p, 
\end{align}
whose Fourier inverse is given by 
\begin{align}
\derpar{}{t} \widetilde{P}(x,v,t)&=\left[ -\derpar{}{x} v + \derpar{}{v}\Gamma(t)(v+v_0) + \dersecpar{}{v}\sigma\Gamma(t) + \frac{\partial^2}{\partial x \partial v}D_{xv}(t)\right]\widetilde{P}(x,v,t) .
\label{FLEkktilde}  
\end{align}
The special case (a) follows straightforwardly, i.e., 
\begin{align}
\derpar{}{t} \widetilde{P}(x,v,t)&=-\derpar{}{x} v \widetilde{P}(x,v,t) + \gamma \derpar{}{v} \left[ (v+v_0) + \sigma\derpar{}{v}\right] \widetilde{P}(x,v,t) .
\end{align}
Clearly, Eqs.~\eqref{FLEkk}, \eqref{FLEkktilde} satisfy property $ii$.

\subsection{L{\'e}vy walk}
\label{silw}

The L{\'e}vy walk model \cite{shlesinger1987levy,sokolov2003towards,zaburdaev2015levy,fedotov2016single} is a special class of the spatiotemporally coupled continuous-time random walk (CTRW) \cite{montroll1965random,fogedby1994langevin,metzler2000random}.
This is typically employed to model position mean-square displacement superdiffusive behaviour, and thus has been widely used to describe transport processes in, e.g., biological systems \cite{zaburdaev2015levy}.  
Here, we study only the $1$-dim case. In the laboratory
frame $\mathcal{S}$ a L\'evy walk is mathematically obtained as
follows: A particle moves with constant speed $u_{\pm}=\pm u$, where
for later convenience we denote by $u_\pm$ its forward/backward velocity, 
for a random running time
$\tau$ sampled by a prescribed distribution $\psi$, after which it
randomly changes its direction of motion. 
The position distribution of a process $X$ performing this type of dynamics is described in terms of master equations, similar to those of the CTRW \cite{metzler2000random}, but with a coupled transition probability
$\phi(y,\tau)=\frac{1}{2}[\delta(y-u_+\tau)+\delta(y-u_-\tau)]\psi(\tau)=\frac{1}{2}\delta(|y|-u \tau)\psi(\tau)$, that relates the walker's position $y$ to the running time $\tau$. 
The GT to the comoving frame $\widetilde{\mathcal{S}}$ expressed by Eq.~\eqref{eq:Gtransf} only changes the walker's velocity as $u_\pm=\pm u -v_0$. 
This is shown easily by transforming $\phi$, which yields
$\widetilde{\phi}(\widetilde{y},\tau)=\frac{1}{2}\delta(|\widetilde{y}+v_0\tau|-u \tau)\psi(\tau)=\frac{1}{2}[\delta(\widetilde{y}+(v_0-u)\tau)+\delta(\widetilde{y}+(v_0+u)\tau)]\psi(\tau)$.
Thus, the microscopic dynamics of L{\'e}vy walks is Galilean
invariant, and we expect its position distribution $P_u$ to correspondingly satisfy weak GI. 
First, we show that property $iii$ is satisfied. 
Remarkably, its position PDF can be obtained exactly in the laboratory frame \cite{zaburdaev2015levy}. 
In fact, denoting $P_0(x)$ the initial distribution and $\Psi(t)=1-\int_0^t \psi(t^{\prime})\diff{t^{\prime}}$ the probability of sampling a running time larger than $t$, $P_u$ is given by   
\begin{equation}
P_u(k,\lambda)=\frac{[\Psi(\lambda - i u k)+\Psi(\lambda + i u k)]P_0(k)}{2-[\psi(\lambda + i u k)+\psi(\lambda - i u k)]} \label{lwPDF} . 
\end{equation}
Identifying in the previous eq. left/right velocities $u_{\pm}$ and substituting for those in the comoving frame $\widetilde{\mathcal{S}}$, 
we obtain the PDF   
\begin{align}
\widetilde{P}_u(k,\lambda)&=\frac{[\Psi(\lambda - i u_+ k)+\Psi(\lambda - i u_- k)]P_0(k)}{2-[\psi(\lambda - i u_+ k)+\psi(\lambda - i u_- k)]} \notag\\
&=\frac{[\Psi(\lambda + i v_0 k - i u k)+\Psi(\lambda + i v_0 k + i u k)]P_0(k)}{2-[\psi(\lambda + i v_0 k - i u k)+\psi(\lambda + i v_0 k + i u k)]}
=P(k,\lambda + i v_0 k) , 
\end{align}
highlighting that the property Eq.~\eqref{newtrans2} 
holds for L{\'e}vy walks ($P$, $\widetilde{P}$ are related by the Laplace variable change $\lambda \to \lambda + i v_0 k$).  

Secondly, we show that property $ii$ also holds. A FP type equation has recently been proposed for L{\'e}vy walks, that has the form in the laboratory frame $\mathcal{S}$ \cite{fedotov2016single}   
\begin{multline}
\left[\dersecpar{}{t}-u^2 \dersecpar{}{x}\right] P_u(x,t)= - \frac{1}{2} \left[ \derpar{}{t} - u \derpar{}{x} \right] \int_0^t K(t^{\prime}) P_u(x- u t^{\prime},t-t^{\prime}) \diff{t^{\prime}} 
- \frac{1}{2} \left[ \derpar{}{t} + u \derpar{}{x} \right]  \int_0^t K(t^{\prime}) P_u(x+u t^{\prime},t-t^{\prime}) \diff{t^{\prime}} ,  
\label{LWFPE}
\end{multline}
with the memory kernel being defined as $K(\lambda)=\psi(\lambda)/\Psi(\lambda)$. It is easy to verify that Eq.~\eqref{LWFPE} yields Eq.~\eqref{lwPDF} in Fourier-Laplace space. 
This equation can be conveniently cast into the form
\begin{equation}
\left[ \dersecpar{}{t}-u^2 \dersecpar{}{x} + \frac{1}{2} \mathcal{D}_t^{(-u,u)} + \frac{1}{2} \mathcal{D}_t^{(u,-u)}\right] P_u(x,t)=0 , 
\end{equation}
where $\mathcal{D}_t^{(v_1,v_2)}$ is the fractional operator 
\begin{equation}
\mathcal{D}^{(v_1,v_2)}_t P_u(x,t)=\left[ \frac{\partial}{\partial t}+v_1\frac{\partial}{\partial x}\right]\int_0^t K(t-t^{\prime})P_u(x-v_2(t-t^{\prime}),t^{\prime}) {\rm d}t^{\prime}, 
\label{LWfracop}
\end{equation}
with Fourier-Laplace representation 
$\mathcal{D}^{(v_1,v_2)}_t
P_u(x,t) \to (\lambda - i k v_1)K(\lambda - i k v_2)P_u(k,\lambda)$. 
For $v_1=v_2=-v_0$, $\mathcal{D}^{(v_1,v_2)}_t$ recovers the fractional substantial derivative Eq.~\eqref{eq:subDeriv}.     
Applying the GT Eq.~\eqref{eq:Gtransf} 
to $\mathcal{D}^{(v_1,v_2)}_t$ in Laplace space yields 
$(\lambda - i k (-v_0+v_1))K(\lambda - i k (-v_0+v_2))P_u(k,\lambda +i k v_0) \to \mathcal{D}^{(-v_0+v_1,-v_0+v_2)}_t \widetilde{P}_u(x,t)$. Therefore, we obtain the FP equation in $\widetilde{\mathcal{S}}$
\begin{equation}
\left[ \dersecpar{}{t}-2v_0 \derpar{}{t}\derpar{}{x}+(v_0^2-u^2) \dersecpar{}{x} + \frac{1}{2} \mathcal{D}_t^{(-v_0-u,-v_0+u)} + \frac{1}{2} \mathcal{D}_t^{(-v_0+u,-v_0-u)}\right] \widetilde{P}_u(x,t)=0 , 
\end{equation}
that can be written more neatly as 
\begin{equation}
\left[ \left( \derpar{}{t}+u_+\derpar{}{x}\right)\left( \derpar{}{t}+u_-\derpar{}{x}\right) + \frac{1}{2} \mathcal{D}_t^{(u_-,u_+)} + \frac{1}{2} \mathcal{D}_t^{(u_+,u_-)}\right] \widetilde{P}_u(x,t)=0 ,  
\end{equation}
which is the correct evolution equation for the L{\'e}vy walk dynamics in the comoving frame $\widetilde{\mathcal{S}}$.  


\section{Derivation of the propagator plotted in Fig.~2A \label{propagator}}

We consider the fractional equation
\begin{equation}
\frac{\partial}{\partial t}\widetilde{P}(x,t)=v_0\frac{\partial}{\partial
  x}\widetilde{P}(x,t)+\sigma\dersecpar{}{x} \mathbb{D}_t \widetilde{P}(x,t),
  \label{mkfracad} 
\end{equation}
where $\mathbb{D}_t$ is the Riemann-Liouville operator with Fourier-Laplace representation $\mathbb{D}_t \widetilde{P}(x,t)\to \lambda^{1-\alpha} \widetilde{P}(k,\lambda)$ ($0<\alpha<1$), that is plotted in Fig.~\ref{mkpdf}a.
Without loss of generality, we assume null initial condition. 
First, we solve Eq.~\eqref{mkfracad} in Fourier-Laplace space:  
\begin{align}
\widetilde{P}(k,\lambda)&=\frac{1}{\lambda^{\alpha^{\prime}}+b(k)\lambda^{\beta}+c(k)}, 
\label{eq:P1Lap}
\end{align}
with the auxiliary parameters $\alpha^{\prime}=1$, $\beta=1-\alpha$, $b(k)=\sigma k^2$ and $c(k)=i k v_0$. Note that $\alpha^{\prime}>\beta$, $\forall \alpha \in (0,1)$. 
We then expand in series as \cite{podlubny1998fractional} 
\begin{align} 
\widetilde{P}(k,\lambda)=\frac{1}{c(k)}\frac{1}{1+\frac{\lambda^{\alpha^{\prime}}+b(k)\lambda^{\beta}}{c(k)}}&=\frac{1}{c(k)}\frac{\lambda^{-\beta}c(k)}{\lambda^{\alpha^{\prime}-\beta}+b(k)}\frac{1}{1+\frac{\lambda^{-\beta}c(k)}{\lambda^{\alpha^{\prime}-\beta}+b(k)}} \notag\\
&=\frac{\lambda^{-\beta}}{\lambda^{\alpha^{\prime}-\beta}+b(k)}\sum_{n=0}^{\infty}(-1)^n \frac{\lambda^{-\beta\,n}[c(k)]^n}{[\lambda^{\alpha^{\prime}-\beta}+b(k)]^n} 
=\sum_{n=0}^{\infty}[-c(k)]^n\frac{\lambda^{-\beta-\beta\,n}}{[\lambda^{\alpha^{\prime}-\beta}+b(k)]^{n+1}} .
\label{eq:seriesE} 
\end{align}
We can now make a term by term Laplace inverse transform of Eq.~\eqref{eq:seriesE} by recalling the formula for the Laplace transform of the three-parameter Mittag-Leffler function given in Eq.~(G14)
. Thus, $\widetilde{P}(k,t)$ is given as  
\begin{align}
\widetilde{P}(k,t)&=\sum_{n=0}^{\infty}[-c(k)]^n t^n \gML{\alpha}{1+n}{1+n}{-t^{\alpha} b(k)} 
=\sum_{n=0}^{\infty}(-i v_0 t)^n k^n \gML{\alpha}{1+n}{1+n}{-\sigma  t^{\alpha} k^2}.
\label{eq:series3ML} 
\end{align}
We now need to make a term by term inverse Fourier transform of Eq.~\eqref{eq:series3ML}. To this aim, we first rewrite it in terms of Fox H-functions by using the corresponding property given in Eq.~(G15).
In our case, we obtain: 
\begin{equation}
\gML{\alpha}{1+n}{1+n}{-\sigma t^{\alpha} k^2}=\frac{1}{\Gamma(1+n)}
H_{1,2}^{1,1}\left[\sigma t^{\alpha} k^2
\left|
\begin{array}{l} 
(-n,1) \\ [0.1cm]
(0,1),(-n,\alpha)
\end{array}
\right.\right].
\end{equation}
Using this formula, the Fourier inverse transform of $k^n \gML{\alpha}{1+n}{1+n}{-\sigma t^{\alpha} k^2}$ is expressed by cosine and sine transforms of Fox H-functions, i.e., it is given by    
\begin{equation}
\frac{1}{2 \pi} \int_{-\infty}^{\infty} \cos{(k  x)} k^n 
H_{1,2}^{1,1}\!\left[\sigma  t^{\alpha} k^2
\left|
\begin{array}{l} 
(-n,1) \\ [0.1cm]
(0,1),(-n,\alpha)
\end{array}
\right.\!\!\right]
\diff{k} 
- \frac{i}{2 \pi} \int_{-\infty}^{\infty} \sin{(k  x)} k^n 
H_{1,2}^{1,1}\left[\sigma  t^{\alpha} k^2
\left|
\begin{array}{l} 
(-n,1) \\ [0.1cm]
(0,1),(-n,\alpha)
\end{array}
\right.\right]
\diff{k}.
\label{eq:FnthT}
\end{equation}

Let us first assume $x>0$. We remark that (a) the first/second integral in Eq.~\eqref{eq:FnthT} is not null only for even/odd indices, i.e., for $n=2 \nu$/$1+2 \nu$, $\forall \nu \in \mathbb{N}_0$ respectively, due to the parity of the Fox H-function, and that (b) they are equal to twice the corresponding integral on the semi-half positive line, once not null. Thus, we can use the property of the H-function given in Eqs.~\eqref{eq:FHCStransf} to compute these integrals:
\begin{subequations}
\begin{align}
\int_{0}^{\infty} \cos{(k  x)} k^{2 \nu} 
H_{1,2}^{1,1}\!\left[\sigma  t^{\alpha} k^2
\left|
\begin{array}{l} 
(-2 \nu,1) \\ [0.1cm]
(0,1),(-2 \nu,\alpha)
\end{array}
\right.\!\!\right]
\diff{k} &= 
\frac{\sqrt{\pi}  2^{2 \nu}}{|x|^{1+2 \nu}}
H_{3,2}^{1,2}\!\left[\frac{4  \sigma   t^{\alpha}}{x^2}
\left|
\begin{array}{l} 
\left(\frac{1}{2}-\nu,1\right), (-2 \nu,1), (-\nu,1) \\ [0.1cm]
(0,1), (-2 \nu,\alpha)
\end{array}
\right.\!\!\right], 
 \\
\int_{0}^{\infty} \sin{(k  x)} k^{1+2 \nu} 
H_{1,2}^{1,1}\!\left[\sigma  t^{\alpha} k^2
\left|
\begin{array}{l} 
(-1-2 \nu,1) \\ [0.1cm]
(0,1),(-1-2 \nu,\alpha)
\end{array}
\right.\!\!\right]
\!\diff{k} &= 
\frac{\sqrt{\pi} 2^{1+2 \nu}}{|x|^{2+2 \nu}}
H_{3,2}^{1,2}\!\left[\frac{4  \sigma   t^{\alpha}}{x^2}
\left|
\begin{array}{l} 
\left(-\frac{1}{2}-\nu,1\right), (-1-2 \nu,1), (-\nu,1) \\ [0.1cm]
(0,1), (-1-2 \nu,\alpha)
\end{array}
\right.\!\!\right].
\end{align}
\end{subequations}  
By using the further property in Eq.~\eqref{eq:invertFH} we obtain: 
\begin{multline}
H_{1,2}^{1,1}\left[\sigma  t^{\alpha} k^2
\left|
\begin{array}{l} 
(-n,1) \\ [0.1cm]
(0,1),(-n,\alpha)
\end{array}
\right.\right] 
\xrightarrow[\text{inverse}]{\text{Fourier}}
\frac{1}{\sqrt{\pi}}\left\{
\begin{array}{ll}
\frac{2^{2 \nu}}{|x|^{1+2 \nu}} 
H_{2,3}^{2,1}\left[\frac{x^2}{4  \sigma   t^{\alpha}}
\left|
\begin{array}{l} 
(1,1), (1+2 \nu,\alpha) \\ [0.1cm]
\left(\frac{1}{2}+\nu,1\right), (1+2 \nu,1), (1+\nu,1)
\end{array}
\right.\right]
& \quad n=2 \nu \vspace{0.2cm}\\
\frac{(- i) 2^{1+2 \nu}}{|x|^{2+2 \nu}}
H_{2,3}^{2,1}\left[\frac{x^2}{4  \sigma   t^{\alpha}}
\left|
\begin{array}{l} 
(1,1), (2+2 \nu, \alpha) \\ [0.1cm]
\left(\frac{3}{2}+\nu,1\right), (2+2 \nu,1), (1+\nu,1)
\end{array}
\right.\right]
& \quad n=1+2 \nu
\end{array}
\right.
\label{eq:nthT}
\end{multline}
These results enable us to write Eq.~\eqref{eq:P1Lap} explicitly in $(x,t)$-space in terms of two infinite series of Fox H-functions (corresponding to the original series over odd and even indices): 
\begin{multline}
\widetilde{P}(x,t)=\frac{1}{\sqrt{\pi}}\sum_{\nu=0}^{\infty} \frac{(-1)^{\nu}(v_0  t)^{2 \nu}}{(2  \nu)!}
\frac{2^{2 \nu}}{|x|^{1+2 \nu}} 
H_{2,3}^{2,1}\!\left[\frac{x^2}{4  \sigma   t^{\alpha}}
\left|
\begin{array}{l} 
(1,1), (1+2 \nu,\alpha) \\ [0.1cm]
\left(\frac{1}{2}+\nu,1\right), (1+2 \nu,1), (1+\nu,1)
\end{array}
\right.\!\!\right] \\
+\!\frac{1}{\sqrt{\pi}}\sum_{\nu=0}^{\infty} \frac{(-1)^{1+\nu}(v_0  t)^{1+2 \nu}}{(1+2 \nu)!}
\frac{2^{1+2 \nu}}{|x|^{2+2 \nu}}
H_{2,3}^{2,1}\!\left[\frac{x^2}{4  \sigma   t^{\alpha}}\left|
\begin{array}{l} 
(1,1), (2+2 \nu, \alpha) \\ [0.1cm]
\left(\frac{3}{2}+\nu,1\right), (2+2 \nu,1), (1+\nu,1)
\end{array}
\right.\!\!\right] .
\label{eq:P1series}
\end{multline}
Finally, we can exploit Eq.~\eqref{eq:absorbFH} to absorb the $x$-dependent multiplicative factors into the Fox H-functions. For each term separately, we obtain: 
\begin{subequations}
\begin{align}
\frac{2^{2 \nu}}{|x|^{1+2 \nu}} 
H_{2,3}^{2,1}\left[\frac{x^2}{4  \sigma   t^{\alpha}}
\left|
\begin{array}{l} 
(1,1), (1+2  \nu,\alpha) \\ [0.1cm]
\left(\frac{1}{2}+\nu,1\right), (1+2  \nu,1), (1+\nu,1)
\end{array}
\right.\right] &= 
\frac{(\sigma   t^{\alpha})^{-\nu}}{\sqrt{4  \sigma   t^{\alpha}}}
H_{2,3}^{2,1}\left[\frac{x^2}{4  \sigma   t^{\alpha}}
\left|
\begin{array}{l} 
\left(\frac{1}{2}-\nu,1\right), \left(1+2 \nu-\alpha\left(\frac{1}{2}+\nu \right),\alpha\right) \\ [0.1cm]
(0,1), \left(\frac{1}{2}+\nu,1\right), \left(\frac{1}{2},1\right)
\end{array}
\right.\right] 
\label{eq:seriesT1} \\
\frac{2^{1+2 \nu}}{|x|^{2+2 \nu}}
H_{2,3}^{2,1}\left[\frac{x^2}{4  \sigma   t^{\alpha}}
\left|
\begin{array}{l} 
(1,1), (2+2 \nu, \alpha) \\ [0.1cm]
\left(\frac{3}{2}+\nu,1\right), (2+2 \nu,1), (1+\nu,1)
\end{array}
\right.\right] &= 
\frac{(\sigma   t^{\alpha})^{-\nu-1/2}}{\sqrt{4  \sigma   t^{\alpha}}}
H_{2,3}^{2,1}\left[\frac{x^2}{4  \sigma   t^{\alpha}}
\left|
\begin{array}{l} 
(-\nu,1), (2+2 \nu-\alpha(1+\nu),\alpha) \\ [0.1cm]
\left(\frac{1}{2},1\right), \left(1+\nu,1\right), (0,1)
\end{array}
\right.\right] .
\label{eq:seriesT2}
\end{align}
\end{subequations}
In the opposite case $x<0$ the second term in the rhs of Eq.~\eqref{eq:FnthT} changes sign, so that the sum over odd indices in Eq.~\eqref{eq:P1series} has an opposite sign as well. 
If we take this into account and substitute Eqs.~\eqref{eq:seriesT1}, \eqref{eq:seriesT2} into Eq.~\eqref{eq:P1series}, we obtain that  
$\widetilde{P}(x,t)$ is defined as an infinite series of Fox H-functions ($\forall x \neq 0$), i.e., 
\begin{equation}
\widetilde{P}(x,t)=\frac{1}{\sqrt{4 \pi \sigma  t^{\alpha}}}\left[ \Theta(x)
\sum_{n=0}^{\infty}\frac{(-1)^n}{n!}\left(\frac{v_0  t}{\sqrt{\sigma   t^{\alpha}}}\right)^n \overline{H}_{2,3}^{2,1}\left(\frac{x^2}{4  \sigma t^{\alpha}}; \alpha, n\right) 
+\Theta(-x) \sum_{n=0}^{\infty}\frac{1}{n!}\left(\frac{v_0  t}{\sqrt{\sigma   t^{\alpha}}}\right)^n \overline{H}_{2,3}^{2,1}\left(\frac{x^2}{4  \sigma t^{\alpha}}; \alpha, n\right)\right],
\label{eq:MKPDFb} 
\end{equation}
where the auxiliary function $\overline{H}_{2,3}^{2,1}\left(x; \alpha, n\right)$ is defined as
\begin{align}
\overline{H}_{2,3}^{2,1}\!\left(x; \alpha,n \right)&=\left\{
\begin{array}{ll}
(-1)^{\nu}
H_{2,3}^{2,1}\!\left[x
\left| 
\begin{array}{l} 
\left(\frac{1-2 \nu}{2},1\right), \left( \frac{\left(2-\alpha\right)(1+2 \nu)}{2}, \alpha \right) \\ [0.1cm]
(0,1), \left(\frac{1+2 \nu}{2},1 \right), \left(\frac{1}{2},1\right)
\end{array}
\right.\!\!\right] & \quad n=2 \nu \vspace{0.1cm} \\ 
(-1)^{\nu}
H_{2,3}^{2,1}\!\left[x
\left| 
\begin{array}{l} 
\left(- \nu,1\right), \left( (2-\alpha)(1+\nu), \alpha \right) \\ [0.1cm]
\left(\frac{1}{2},1\right), \left(1+\nu,1\right), (0,1)
\end{array}
\right.\!\!\right] & \quad n=1+2 \nu
\end{array}
\right.
\label{eq:Hbar}
\end{align}
The previous formula is valid for $x\neq0$. Therefore, we need to specify the value of the PDF in this point. In this case, only the sum over even indices contributes to the PDF in Eq.~\eqref{eq:series3ML} (the sine transform in Eq.~\eqref{eq:FnthT} is, in fact, null) with coefficients defined by solving the correspondent integral of Fox function with Eqs.~\eqref{eq:thetaFH}, (G11):
\begin{equation}
\int_{0}^{\infty} k^{2 \nu} H_{1,1}^{1,2}\left[\sqrt{\sigma  t^{\alpha}} |k|
\left|
\begin{array}{l} 
\left(-2 \nu,\frac{1}{2}\right) \\ [0.1cm]
\left(0,\frac{1}{2}\right), \left(-2 \nu,\frac{\alpha}{2}\right)
\end{array}
\right.\right]\diff{k}
=\left(\frac{1}{\sqrt{\sigma  t^{\alpha}}}\right)^{1+2 \nu}\frac{\left[\Gamma\left(\frac{1}{2}+\nu\right)\right]^2}{\Gamma\left((1+2 \nu)\left(1-\frac{\alpha}{2}\right)\right)}. 
\label{eq:1.3.10}
\end{equation}
By substituting such coefficients into the series over even indices, we obtain:
\begin{align}
\widetilde{P}(0,t)&=\frac{1}{\sqrt{4 \sigma  t^{\alpha}}}\sum_{\nu=0}^{\infty}\frac{(-1)^{\nu}(2 \nu)!}{4^{\nu}(\nu !)^2}\frac{1}{\Gamma\left(\left(1-\frac{\alpha}{2}\right)(1+2 \nu)\right)}\left(\frac{v_0^2 t^{2-\alpha}}{4 \sigma }\right)^{\nu} \notag\\
&=\frac{1}{\sqrt{4 \sigma  t^{\alpha}}}\gML{2-\alpha}{(2-\alpha)/2}{1/2}{-\frac{v_0^2 t^{2-\alpha}}{4 \sigma }} . 
\end{align} 

Note that Eq.~\eqref{eq:MKPDFb} is expressed as an expansion in the constant force field $v_0$, i.e., the velocity of the frame $\widetilde{\mathcal{S}}$. As a sanity check, we compute the zero-th order term, which must be equal to the solution in the frame $\mathcal{S}$, i.e., the position PDF of a force-free CTRW \cite{metzler2000random}. This is confirmed below (note that the corresponding terms in the two series in Eq.~\eqref{eq:MKPDFb} are equal):
\begin{align}
\widetilde{P}(x,t)&=\frac{1}{\sqrt{4 \pi \sigma  t^{\alpha}}}
H_{2,3}^{2,1}\!\left[ \frac{x^2}{4 \sigma t^{\alpha}}
\left|
\begin{array}{l} 
\left(\frac{1}{2},1\right), \left( \frac{2-\alpha}{2}, \alpha \right) \\ [0.1cm]
\left(0,1\right),\left(\frac{1}{2},1\right), \left(\frac{1}{2},1\right)
\end{array}
\right.\!\!\right] 
=\frac{1}{\sqrt{4 \pi \sigma  t^{\alpha}}}
H_{1,2}^{2,0}\!\left[ \frac{x^2}{4 \sigma t^{\alpha}}
\left|
\begin{array}{l} 
\left( \frac{2-\alpha}{2}, \alpha \right) \\ [0.1cm]
\left(0,1\right),\left(\frac{1}{2},1\right)
\end{array}
\right.\!\!\right] .
\label{eq:zerothTerm}
\end{align}
Here, we used the property of the Fox H-function given in Eq.~\eqref{eq:reduction1}. 

At last, we check the normalisation of the derived formula for $\widetilde{P}$, which is expected as $\widetilde{P}(k=0,\lambda)=1/\lambda$. Due to the different sign of the sums over odd indices, only those over even ones contribute to the normalization of the PDF. Due to the parity of the Fox H-function, the integral can be restricted to the semi-half positive line:  
\begin{equation}
\int_{-\infty}^{\infty} \widetilde{P}(x,t)\diff{x}=\frac{1}{\sqrt{\pi \sigma  t^{\alpha}}}\sum_{\nu=0}^{\infty}\frac{(-1)^{\nu}}{(2 \nu)!}\left(\frac{v_0 t}{\sqrt{\sigma   t^{\alpha}}}\right)^{2 \nu} 
\int_{0}^{\infty} H_{2,3}^{2,1}\!\left[\frac{x^2}{4 \sigma  t^{\alpha}}
\left|
\begin{array}{l} 
\left(\frac{1-2 \nu}{2},1\right), \left( (2-\alpha)\left(\frac{1+2 \nu}{2}\right), \alpha \right) \\ [0.1cm]
\left(0,1\right), \left(\frac{1+2 \nu}{2},1 \right), \left(\frac{1}{2},1\right)
\end{array}
\right.\right] \diff{x}.
\label{eq:normalization}
\end{equation}
We compute the integral of the Fox H-function by recalling Eqs.~\eqref{eq:scalingFH}, (G11): 
\begin{align}
\int_{0}^{\infty} H_{2,3}^{2,1}\!\left[\frac{x^2}{4 \sigma  t^{\alpha}}
\left|
\begin{array}{l} 
\left(\frac{1-2 \nu}{2},1\right), \left( (2-\alpha)\left(\frac{1+2 \nu}{2}\right), \alpha \right) \\ [0.1cm]
\left(0,1\right), \left(\frac{1+2 \nu}{2},1 \right), \left(\frac{1}{2},1\right)
\end{array}
\right.\right] \diff{x}&=
\sqrt{\sigma  t^{\alpha}}\Theta(-1),
\label{eq:evenI}
\end{align}
where the function $\Theta$ is defined in Eq.~\eqref{eq:thetaFH}, which in this specific case is  
\begin{align}
\Theta(s)\!&=\!
\frac{\Gamma\!\left(\frac{s}{2}\right)\Gamma\!\left(\frac{1}{2}+\nu + \frac{1}{2}s\right)\Gamma\!\left(\frac{1}{2}-\frac{s}{2}+\nu \right)}{\Gamma\!\left(\frac{1}{2}-\frac{1}{2}s\right)\Gamma\!\left((2-\alpha)\left(\frac{1+2 \nu}{2}\right)+\frac{\alpha}{2}s \right)}\!=\! 
\frac{\Gamma\!\left(\frac{s}{2}\right)\Gamma\!\left(\frac{1}{2}+\nu + \frac{1}{2}s\right)}{\Gamma\!\left((2-\alpha)\left(\frac{1+2 \nu}{2}\right)+\frac{\alpha}{2}s \right)}\prod_{i=0}^{\nu-1}\left(\frac{1}{2}-\frac{s}{2}+i \right).
\end{align}
For $s\!=\!-1$ all terms, except that for $\nu=0$, which is equal to $\sqrt{\pi}$, cancel out. Eq.~\eqref{eq:evenI} is then equal to $\sqrt{\pi \sigma  t^{\alpha}}$, i.e., the PDF is correctly normalised.

\section{Derivation of the characteristic functional of the noise $\overline{\xi}$ \label{xibar}}

The noise $\overline{\xi}$ can be formally defined as \cite{cairoli2015langevin}
\begin{equation}
\overline{\xi}(t)=\int_0^{\infty}\xi(s)\delta(t-T(s))\diff{s} ,
\label{xiBintdef} 
\end{equation}
where $\xi$ is a white Gaussian noise with $\langle \xi(t)\rangle=0$ and $\langle \xi(t_1)\xi(t_2)\rangle=2\sigma\delta(t_1-t_2)$, and $T$ is a strictly increasing L{\'e}vy process \cite{cont1975financial}. 
Within the subordination description of CTRWs \cite{fogedby1994langevin,magdziarz2009langevin,cairoli2015anomalous,cairoli2017timedep}, they specify respectively the stochastic process of jump lengths and that of waiting times of the underlying random walk. 
We recall the definition of the inverse subordinator     
$S(t)=\inf_{s>0}{\left\{s:T(s)>t\right\}}$,
such that 
$\int_0^t \diff{t^{\prime}} \overline{\xi}(t^{\prime})=B(S(t))$, where $B$ is an ordinary Brownian motion. 

Its characteristic functional is defined for a general test function $u(r)$ as 
\begin{equation}
G[u(r)]=\left\langle \exp{\left( i \int_0^{\infty} u(r)\,\overline{\xi}(r)\diff{r}\right)} \right\rangle.
\label{eq:charF}
\end{equation} 
Note that the brackets denote an average over the realisations of both the stochastic processes $\xi$ and $T$ specifying Eq.~\eqref{xiBintdef}.    
By substituting this definition into Eq.~\eqref{eq:charF}, we obtain
\begin{align}
G[u(s_1)]&=\Average{\exp{\left[i \int_0^{\infty} u(s_1)\left(\int_0^{\infty}\xi(s_2)\delta(s_1-T(s_2))\diff{s_2}\right)\diff{s_1}\right]}} \notag\\
&=\Average{\exp{\left[ i \int_0^{\infty} \xi(s_2)\left( \int_0^{\infty}u(s_1)\delta(s_1-T(s_2))\diff{s_1}\right)\diff{s_2}\right]}} \notag\\
&=\Average{\exp{\left[ i \int_0^{\infty} \xi(s_1)\,f(s_1) \diff{s_1}\right]}} .
\label{eq:xiBChF1}
\end{align}
In the previous expression, we changed the order of integration and defined the auxiliary function  
\begin{equation}
f(s)=\int_0^{\infty}u(s^{\prime})\delta(s^{\prime}-T(s))\diff{s^{\prime}}, 
\label{eq:etaHtest}
\end{equation}
which depends only on the different realisations of the process $T$. For each of them, $f$ is completely determined and it can be used as a test function in the characteristic functional of $\xi$. 
Thus, Eq.~\eqref{eq:xiBChF1} can be simplified if we compute the average over $\xi$ first.   
For a Gaussian noise of correlation function $\Average{\xi(s_1)\,\xi(s_2)}=C(|s_2-s_1|)$, we obtain \cite{feynman2010quantum} 
\begin{equation}
\Average{\exp{\left( i \int_0^{\infty} \xi(s)\,f(s) \diff{s}\right)}}=\Average{\exp{\left( -\frac{1}{2}\int_0^{\infty}\int_0^{\infty} f(s_1)f(s_2)C(s_2-s_1)\diff{s_1}\diff{s_2}\right)}}. 
\label{eq:xiBChF2}
\end{equation}  
The remaining average in its rhs is only on the realizations of the L{\'e}vy process $T$. 
Substituting Eq.~\eqref{eq:etaHtest} into Eq.~\eqref{eq:xiBChF2} yields  
\begin{subequations}
\begin{align}
G[u(r)]&=\Average{\exp{\left( -\int_0^{\infty} \int_0^{\infty} u(r_1)u(r_2)\Lambda(r_1,\,r_2;\,T)\diff{r_1}\diff{r_2}\right)}}, \label{eq:xiBChFA} \\
\Lambda(r_1,\,r_2;\,T)&=\frac{1}{2}\int_0^{\infty}\int_0^{\infty}\delta(r_1-T(s_1))\delta(r_2-T(s_2))C(s_2-s_1)\diff{s_1}\diff{s_2}.
\label{eq:xiBChFB}
\end{align}
\end{subequations}
For $\xi$ white noise with correlation function $C(s_2-s_1)=2\sigma\delta(s_2-s_1)$, Eq.~\eqref{eq:xiBChFB} reduces to   
\begin{align}
\Lambda(r_1,\,r_2;\,T)&=\sigma\int_0^{\infty}\int_0^{\infty} \delta(r_1 -T(s_1)) \delta(r_2 -T(s_2)) \delta(s_2-s_1) \diff{s_1}\diff{s_2} \notag\\
&=\sigma\int_0^{\infty} \delta(r_1 -T(s)) \delta(r_2 -T(s)) \diff{s} 
\notag\\ &
= \sigma \delta(r_2 - r_1) \int_0^{\infty} \delta(r_1 -T(s)) \diff{s} .
\end{align}
Substituting this result into Eq.~\eqref{eq:xiBChFA}, we obtain the characteristic functional, i.e.,  
\begin{align}
G[u(r)]&=\Average{\exp{\left[ -\sigma \int_0^{\infty} \int_0^{\infty} \int_0^{\infty} u(r_1)u(r_2) 
\delta(r_2 - r_1) \delta(r_1 -T(s)) \diff{s} \diff{r_1}\diff{r_2}\right]}} \notag\\
&=\Average{\exp{\left[ -\sigma \int_0^{\infty} \int_0^{\infty} [u(r)]^2 \delta(r -T(s)) \diff{s} \diff{r}\right]}} 
\notag\\&
=\Average{\exp{\left[ -\sigma \int_0^{\infty} [u(T(s))]^2 \diff{s}\right]}} .
\end{align} 
As a sanity check, we calculate the PDF of the process $Y$, satisfying the LE  $\dot{Y}(t)=\overline{\xi}(t)$. 
If we set $u(r)=k \Theta(t-r)$ and employ the relation 
$\Theta(t-T(s))=1-\Theta(s-S(t))$ \cite{baule2005joint}, we find  
\begin{align}
P(k,t)&=\left\langle \exp{\left( -\sigma k^2 \int_0^{\infty} \Theta(t-T(s)) \diff{s} \right)} \right\rangle 
=\left\langle \exp{\left(-\sigma k^2 S(t)\right)} \right\rangle ,  
\end{align}
which is the correct position PDF of a free diffusive CTRW \cite{fogedby1994langevin}. 

Similarly, we can use this technique to prove Eq.~\eqref{fracad} 
and find its propagator. For simplicity, we set the initial condition $Y_0=0$. Recalling that the PDF of the process $\widetilde{Y}$ satisfying the LE 
$\dot{\widetilde{Y}}(t)=-v_0 + \overline{\xi}(t)$
is  
$\widetilde{P}(k,t)=\langle \exp{[i k (-v_0 t + \int_0^t \overline{\xi}(s)\diff{s})]} \rangle$, we can write: 
\begin{equation}
e^{i k v_0 t} \widetilde{P}(k,t)=\Average{e^{-\sigma k^2 S(t)}}=\int_0^{\infty} h(s,t) e^{-\sigma k^2 s} \diff{s}, 
\end{equation}
where $h(s,t)=\langle \delta(s-S(t))\rangle$ \cite{magdziarz2009langevin,cairoli2015anomalous} is the PDF of the inverse subordinator $S$. Then, we first take its time derivative, i.e., 
\begin{equation}
\left[ i k v_0 + \derpar{}{t} \right] \widetilde{P}(k,t) =e^{-i k v_0 t} \int_0^{\infty} e^{-\sigma k^2 s} \derpar{}{t} h(s,t)  \diff{s}, 
\end{equation}
and secondly its Laplace transform. 
Recalling that $\widetilde{h}(s,\lambda)=[\Phi(\lambda)/\lambda] e^{-s \Phi(\lambda)}$, we obtain 
\begin{equation}
\lambda \widetilde{P}(k,\lambda)-1=-i k v_0 \widetilde{P}(k,\lambda)-\sigma k^2 \frac{\lambda +i v_0 k}{\Phi(\lambda +i v_0 k)}\widetilde{P}(k,\lambda), 
\end{equation}  
which is the Laplace transform of Eqs.~\eqref{fracad}, \eqref{eq:subDeriv}. 
Further solving it for $\widetilde{P}$, yields the propagator  
\begin{equation}
\widetilde{P}(k,\lambda)=\frac{1}{\lambda + i k v_0} \left[ 1- \frac{\sigma k^2}{\Phi(\lambda + i k v_0)+\sigma k^2} \right]\:.
\label{eq:SI-PDFcostF}
\end{equation}
For the particular case of $T$ being a L{\'e}vy stable process of order $\alpha$, its inverse Laplace transform is 
\begin{equation}
\widetilde{P}(x,t)=\frac{1}{\sqrt{4 \sigma t^{\alpha}}} H_{1,1}^{1,0}\!
\left[\frac{|x-v_0 t|}{\sqrt{\sigma t^{\alpha}}}\left| \!\!
\begin{array}{c} 
\left(1-\frac{\alpha}{2},\frac{\alpha}{2}\right) \\ [0.1cm]
(0,1)
\end{array}
\right.\!\!\right] ,
\label{eq:GIPDF} 
\end{equation}
which is the PDF plotted in Fig.~\ref{mkpdf}b.

\section{Derivation of the nonlocal advection-diffusion equation 13 via subordination \label{subordination}}

A CTRW is mathematically defined by a normal diffusive process $X$ and a strictly increasing L{\'e}vy process $T$ respectively specifying the
stochastic process of jump lengths and that of waiting times of the
random walk underlying its dynamics in the continuum limit \cite{fogedby1994langevin,magdziarz2009langevin,cairoli2015anomalous}. 
Their dynamics is described by the LEs 
\begin{align}
\dot{X}(s)&=\xi(s), & \dot{T}(s)&=\eta(s) ,
\label{eq:fogedby}
\end{align}
where $\xi$ is Gaussian white noise with $\left< \xi(s)\right>=0$
and $\left< \xi(s_1)\xi(s_2)\right>=2\sigma\delta(s_2 -
s_1)\:,\:\sigma>0$, and $\eta$ is a one-sided positive L{\'e}vy
process with characteristic functional \cite{magdziarz2009langevin,cairoli2015anomalous,cairoli2017timedep}
\begin{equation}
G[u(r)]=\left< e^{ -\int_{0}^{\infty}u(r)\eta(r)\diff{r}}\right>=e^{ -\int_{0}^{\infty}\Phi(u(r))\diff{r}}\:,
\label{eq:LevyChF}
\end{equation}
where $u$ is an arbitrary test function. The function $\Phi$ is the
Laplace exponent of $\eta$ and is in general a Bernstein function \cite{schilling2012bernstein}. 
The anomalous CTRW process $Y$ is
defined by subordination of $X$ with the inverse of $T$, i.e.,
$Y(t)=X(S(t))$, where $S$ is the first passage time process
$S(t)=\inf_{s>0}{\left\{s:T(s)>t\right\}}$.  
In the special case
$\Phi(\lambda)=\lambda^{\alpha}\:,\: 0<\alpha<1$,
Eq.~\eqref{eq:LevyChF} specifies a L{\'e}vy stable process that yields
a subdiffusive CTRW with mean-square displacement that scales for long times as $t^\alpha$.

A similar description can be defined for the process $\widetilde{Y}(t)$ satisfying Eq.~~\eqref{fracad}
, i.e., we set  
$\widetilde{Y}(t)=\widetilde{X}(S(t))$, where $\widetilde{X}$ is described by the LE 
$\dot{\widetilde{X}}(s)=-v_0 \eta(s)+\xi(s)$ instead of Eq.~\eqref{eq:fogedby}(left).  
As pointed out in the main text, weak GI requires a coupling between the LEs of the jump process $X$ and that of the elapsed time process $T$. Here, we prove that its corresponding FP equation is Eq.~\eqref{fracad}
, following the technique of refs.~\cite{cairoli2015anomalous,cairoli2017timedep}.
The time-change $S$ has continuous stochastic paths, such that $\widetilde{Y}$ is a continuous semi-martingale. Thus, its It{\^o} formula for an arbitrary test function $f$ is 
\begin{equation}
f(\widetilde{Y}(t)) =  f(Y_0) + \int_0^t \derpar{}{y}f(\widetilde{Y}(t^{\prime}))\diff{\widetilde{Y}(t^{\prime})} + \frac{1}{2}\int_0^t \dersecpar{}{y}f(\widetilde{Y}(t^{\prime}))\diff{[\widetilde{Y},\widetilde{Y}]_{t^{\prime}}} , 
\label{eq:ItoFsm} 
\end{equation}  
where $\widetilde{Y}(0)=Y_0$ is the initial condition and $[\widetilde{Y},\widetilde{Y}]_t= 2 \sigma \int_0^t \diff{S(t^{\prime})}$ is its quadratic variation. 
If we now evaluate Eq.~\eqref{eq:ItoFsm} for the specific choice $f(\widetilde{Y}(t))=e^{i k \widetilde{Y}(t)}$, we obtain:
\begin{align}
e^{i k \widetilde{Y}(t)}
&=e^{i k x_0}+i k \int_0^t e^{i k \widetilde{Y}(t^{\prime})}\diff{\widetilde{Y}(t^{\prime})}-\sigma k^2 \int_0^t e^{i k \widetilde{Y}(t^{\prime})}\diff{S(t^{\prime})} \notag \\
&=e^{i k x_0}  - i k v_0 \int_0^t e^{i k \widetilde{Y}(t^{\prime})}\diff{t^{\prime}} + i k \int_0^t e^{i k \widetilde{Y}(t^{\prime})} \xi(S(t^{\prime}))\diff{S(t^{\prime})} -  \sigma k^2 \int_0^t e^{i k \widetilde{Y}(t^{\prime})}\diff{S(t^{\prime})} . 
\label{eq:ItoExp}
\end{align} 
Here, we substituted the stochastic trajectory of $\widetilde{Y}$, obtained by exact integration of its LE. 
Thus, if we now (a) ensemble average Eq.~\eqref{eq:ItoExp} (which cancels out the third term in its rhs because $\xi$ is Gaussian noise with null first moment), (b) make its Fourier inverse transform and (c) take the time derivative of the resulting equation, we obtain: 
\begin{equation}
\derpar{}{t}\widetilde{P}(x,t)= v_0 \derpar{}{x} \widetilde{P}(x,t)+ \sigma \dersecpar{}{x}\derpar{}{t}\left\langle \int_0^t \delta(x-\widetilde{Y}(t^{\prime}))\diff{S(t^{\prime})}\right\rangle.    
\label{eq:FPsm}
\end{equation}
Let us now compute the averaged stochastic integral in its rhs \cite{cairoli2015anomalous,cairoli2017timedep}.  
Employing the relation $1=\int_0^{\infty} \delta(s-S(t))\diff{s}$, we define an auxiliary quantity $Q$ as  
\begin{align}
Q(x,t)&=\left\langle \int_0^t \delta(x-\widetilde{Y}(t^{\prime})) \diff{S(t^{\prime})}\right\rangle 
=\left\langle \int_0^t \left[ \int_0^{\infty} \delta(x-\widetilde{X}(s))\delta(s-S(t^{\prime}))\diff{s}\right] \diff{S(t^{\prime})}\right\rangle ,
\label{eq:Q}
\end{align}
leading in Fourier transform to
\begin{align}
Q(k,t)&=\left\langle \int_0^t \left[ \int_0^{\infty} e^{i k \widetilde{X}(s)}\delta(s-S(t^{\prime}))\diff{s}\right] \diff{S(t^{\prime})\!}\right\rangle \notag\\
&=\int_0^{t} \left[ \int_0^{\infty} \left\langle e^{i k \int_0^s \xi(r)\diff{r}}\right\rangle \left\langle e^{-i k v_0 T(s)}\delta(t^{\prime}-T(s))\right\rangle  \diff{s} \right]  \diff{t^{\prime}} . 
\label{eq:FouQ}
\end{align}
This equation is obtained by recalling that $\Theta(s-S(t))=1-\Theta(t-T(s))$ \cite{baule2005joint}, which, together with the continuity of the paths of $S$, implies the relation: $\delta(t-T(s))=\delta(s-S(t))\dot{S}(t)$ \cite{cairoli2015anomalous,cairoli2017timedep}. Here, $\dot{S}(t)=\lim_{\Delta t \to 0} [S(t+\Delta t)-S(t)]/\Delta t$ denotes an integration with respect to the time-change $S$. 
This is conveniently employed to express the stochastic integral in the left-hand side (lhs) of Eq.~\eqref{eq:FouQ} in terms of time increments. 
By introducing a partition of the interval $[0,t]$ of finite mesh $\Delta t$, we can write ($N=t/\Delta t$):   
\begin{align}
\int_{0}^{t}\left[ \int_0^{\infty} e^{i k \widetilde{X}(s)} \delta(s-S(t^{\prime})) \diff{s} \right] \diff{S(t^{\prime})}
&=\lim_{N \to \infty \atop \Delta t \to 0} \sum_{i=0}^{N-1} \left[ \int_0^{\infty} e^{i k \widetilde{X}(s)}\delta(s-S(t^{\prime}_i))\diff{s}\right] [S(t^{\prime}_{i+1})-S(t^{\prime}_i)] \notag\\
&=\lim_{N \to \infty \atop \Delta t \to 0} \sum_{i=0}^{N-1} \left[ \int_0^{\infty} e^{i k \widetilde{X}(s)}\delta(t^{\prime}_i-T(s)) \diff{s}\right](t^{\prime}_{i+1}-t^{\prime}_i) 
\notag\\
&
=\int_0^t \left[ \int_0^{\infty} e^{i k \widetilde{X}(s)}\delta(t^{\prime} -T(s))\diff{s}\right]\diff{t^{\prime}}.
\label{eq:GFFKstochI}
\end{align}
Eq.~\eqref{eq:FouQ} then follows from Eq.~\eqref{eq:GFFKstochI} by substituting the exact expression of $\widetilde{X}$ and by using the independence of $\xi$ and $\eta$ to factorise the ensemble average. 
Finally, we take the Laplace transform of Eq.~\eqref{eq:FouQ} to obtain:  
\begin{align}
Q(k,\lambda)&=\frac{1}{\lambda}\int_0^{\infty} \left\langle e^{i k \int_0^s \xi(r)\diff{r}}\right\rangle \left\langle e^{-(\lambda + i k v_0)\,T(s)}\right\rangle \diff{s}
=\frac{1}{\lambda}\int_0^{\infty} \left\langle e^{i k \int_0^s \xi(r)\diff{r}}\right\rangle e^{-s \Phi(\lambda + i k v_0)} \diff{s},
\label{eq:LapFouQ}
\end{align}     
where the average over $T$ is computed by employing its characteristic functional Eq.~\eqref{eq:LevyChF}.  

On the other hand, we can rewrite the position PDF of $\widetilde{Y}$ by using (a) the relation with which Eq.~\eqref{eq:Q} has been obtained, (b) the definition of $\widetilde{X}$ and (c) the independence of $\xi$ and $\eta$. We then obtain in Fourier space:
\begin{align}
\widetilde{P}(k,t)&=\int_0^{\infty} \left\langle \delta(s-S(t)) e^{i k \widetilde{X}(s)}\right\rangle \diff{s} =\int_0^{\infty}\left\langle \delta(s-S(t)) e^{-i k v_0 T(s)}\right\rangle \left\langle e^{i k \int_0^s \xi(r)\diff{r}}\right\rangle \diff{s} ,
\end{align}
whose Laplace transform can be calculated by recalling that $\int_0^{\infty} \delta(s-S(t))e^{-\lambda t} \diff{t}=\eta(s) e^{-\lambda T(s)}$ \cite{cairoli2015anomalous}. 
We find:  
\begin{align}
\widetilde{P}(k,\lambda) & = \int_0^{\infty}\left\langle \eta(s) e^{-(\lambda + i k v_0) T(s)}\right\rangle \left\langle e^{i k \int_0^s \xi(r)\diff{r}}\right\rangle \diff{s}.  
\label{eq:LapFouQ2} 
\end{align}
The $\eta$-dependent term can then be rewritten as
\begin{align}
\left\langle \eta(s) e^{-(\lambda + i k v_0) T(s)}\right\rangle 
&=\frac{-1}{\lambda + i k v_0} \dertot{}{s}\left\langle e^{-(\lambda + i k v_0)\int_0^{s} \eta(s^{\prime}) \diff{s^{\prime}}} \right\rangle \notag\\ 
&=\frac{-1}{\lambda + i k v_0} \dertot{}{s} e^{-s \Phi(\lambda + i k v_0)} 
= \frac{\Phi(\lambda + i k v_0)}{(\lambda + i k v_0)} e^{-s \Phi(\lambda + i k v_0)} ,
\label{eq:etaT}
\end{align}
where we used again Eq.~\eqref{eq:LevyChF} with $u(r)=\Theta(s-r)(\lambda + i k v_0)$. Substituting this result into Eq.~\eqref{eq:LapFouQ2}, we obtain:
\begin{equation}
\int_0^{\infty} \left\langle e^{i k \int_0^s \xi(r)\diff{r}}\right\rangle e^{-s \Phi(\lambda + i k v_0)}\diff{s}=\frac{(\lambda + i k v_0)}{\Phi(\lambda + i k v_0)} \widetilde{P}(k,\lambda) .
\label{eq:IntToP}
\end{equation} 
The lhs of Eq.~\eqref{eq:IntToP} coincides with the integral at the rhs of Eq.~\eqref{eq:LapFouQ}. By eliminating it, we obtain  
\begin{equation}
\lambda Q(k,\lambda)=\frac{(\lambda + i k v_0)}{\Phi(\lambda + i k v_0)} \widetilde{P}(k,\lambda),
\end{equation}
or equivalently in $(k,t)$-space (recalling that $Q(x,0)=0$ by definition): 
\begin{equation}
\derpar{}{t}Q(k,t)=\left[ i k v_0 + \derpar{}{t} \right] \int_0^t e^{-i k v_0 (t-s)} K(t -s) \widetilde{P}(k,s) \diff{s} .  
\label{eq:Qrel}
\end{equation} 
Finally, by taking its inverse Fourier transform and substituting it back into Eq.~\eqref{eq:FPsm}, we derive Eq.~\eqref{fracad}.

\section{Derivation of the nonlocal advection-diffusion equation 13 in the superdiffusive regime \label{superdiff}}

We consider the stochastic process $\widetilde{Y}(t)$ in the comoving frame $\widetilde{\mathcal{S}}$, whose dynamics is described by the LE 
$\dot{\widetilde{Y}}(t)\!=\!-v_0+\overline{\xi}(t)$, where 
the noise $\overline{\xi}$ is defined by its hierarchy of correlation functions; specifically, the odd ones are null, i.e., $\langle \prod_{j=1}^{1 + 2 N} \overline{\xi}(t_j) \rangle = 0$, while the even ones are \cite{cairoli2015langevin}
\begin{equation}
\Average{\prod_{j=1}^{2 N} \overline{\xi}(t_j)}  = \frac{\sigma^{N/2}}{N!  2^N } \sum_{\varsigma \in \varSigma_{2 N}}\prod_{m=1}^{N} \delta\!\left(t_{\varsigma(2 N - m + 1)}-t_{\varsigma(m)}\right) 
\sum_{\varsigma^{\prime} \in \varSigma_{N}} \Theta\!\left(t_{\varsigma(\varsigma^{\prime}(m))}-t_{\varsigma(\varsigma^{\prime}(m-1))}\right)\! K\!\left(t_{\varsigma(\varsigma^{\prime}(m))}-t_{\varsigma(\varsigma^{\prime}(m-1))}\right).   
\label{eq:xiBcorrelations}
\end{equation}
Here, $\varsigma$($\varsigma^{\prime}$) is a permutation of $2 N$($N$) elements, which keeps the initial time fixed, $\varSigma_{2 N}$($\varSigma_N$) denotes the set of all such operations, $\Theta$ is an Heaviside function and $K$ an arbitrary function of time. 
Eq.~\eqref{eq:xiBcorrelations} represents an equivalent characterisation of the noise obtained by time derivative of a subordinated Brownian motion \cite{cairoli2015langevin} (appendix~\ref{xibar}), in which case $K$ is related to the Laplace exponent $\Phi$ of a strictly increasing L{\'e}vy process $T$ by the formula $K(\lambda)\!=\!\Phi(\lambda)^{-1}$ \cite{magdziarz2009langevin,cairoli2015anomalous}. 
This generally yields subdiffusive MSD behaviour. 
However, Eq.~\eqref{eq:xiBcorrelations} still characterises a well-defined noise, even if a corresponding process $T$ cannot be defined. 
Thus, $Y$ may exhibit even super-diffusive behaviour, e.g., by setting $K(t)=t^{\alpha-1}/\Gamma(\alpha)$ for $1<\alpha<2$.  

Recalling Eq.~\eqref{eq:GaussCorrFP2}, we need to compute the averaged quantity $\langle \overline{\xi}(t) h(k,t)\rangle$, where we set $h(k,t)=e^{-i k v_0 t + \int_0^t \overline{\xi}(s)\diff{s}}$.  
In the expression of $h$, the second exponential is a functional of the noise path, that can be Taylor expanded as \cite{novikov1965functionals,hanggi1978correlation} 
\begin{align}
e^{i k v_0 t}h(k,t)-1&= \sum_{n=1}^{\infty}\frac{1}{n!}\int_0^{\infty}\diff{s_1}\ldots\int_0^{\infty}\diff{s_n}
H^{(n)}(k,s_1,\ldots ,s_n) \overline{\xi}(s_1)\ldots\overline{\xi}(s_n) \notag\\
& = \sum_{n=1}^{\infty}\frac{(i k)^n}{n!}\int_0^{t}\diff{s_1}\ldots\int_0^{t}\diff{s_n}\overline{\xi}(s_1)\ldots\overline{\xi}(s_n) ,
\label{eq:xiBfuncTaylor}
\end{align}
where the variational derivatives are $H^{(n)}(k,s_1,\ldots ,s_n)=\frac{\delta^{(n)}e^{i k \int_0^t \overline{\xi}(s)\diff{s}}}{\delta\overline{\xi}(s_1)\ldots\delta\overline{\xi}(s_n)}\Big|_{\overline{\xi}=0}=(i k)^n \Theta(t-s_1)\ldots\Theta(t-s_n)$ .

Let us take the ensemble average of Eq.~\eqref{eq:xiBfuncTaylor} and then its time derivative. 
As the odd correlation functions of $\overline{\xi}$ are null, only the terms with even indices survive, so that we obtain:
\begin{multline}
\left[i k v_0 +\derpar{}{t}\right]\widetilde{P}(k,t)=e^{-i k v_0 t} 
\left[\frac{\sigma(i k)^2}{2} K(t) 
+\frac{\sigma^2(i k)^4}{4} \int_0^t K(t-s)K(s)\diff{s} \right. \\
 \left. + \sum_{n=3}^{\infty}\frac{\sigma^n (i k)^{2 n}}{2^n}
\int_0^{t}\diff{s_{n-1}}K(t-s_{n-1})\ldots\int_0^{s_2}K(s_2-s_1)K(s_1)\diff{s_{1}}\right] .
\label{eq:xiBfuncTaylor2}
\end{multline}
This result is understood by recalling that the ($2 n$)-th order correlation function of $\overline{\xi}$ contains $(2 n)!/(2^n n!)$ terms, each corresponding to a different structure of the delta functions. In addition, for each of the sequences of the $n$ distinct times, set by the product of delta functions, there are $n!$ different orderings. However, once we integrate over time, all of them give the same contribution, so that we obtain $(2 n)!/2^n$ integrals of the same type, thus leading to the final result Eq.~\eqref{eq:xiBfuncTaylor2}.      

We then multiply Eq.~\eqref{eq:xiBfuncTaylor} by $\overline{\xi}(t)$ and take its ensemble average. By eliminating the null terms, we obtain:
\begin{align}
\left\langle \overline{\xi}(t) h(k,t) \right\rangle &= 
\sum_{n=0}^{\infty}\frac{\sigma^{1+n}(i k)^{1+2 n}}{(1+2 n)!}e^{-i k v_0 t} 
\int_0^{t}\diff{s_1}\ldots\int_0^{t}\diff{s_{1+2 n}}
\left\langle\overline{\xi}(t)\overline{\xi}(s_1)\ldots\overline{\xi}(s_{1+2 n})
\right\rangle .
\label{eq:xiBF}
\end{align}
We then find  
(a)$\int_0^t \diff{s_1} \left\langle \overline{\xi}(t)\overline{\xi}(s_1)\right\rangle=\sigma K(t)$ for $n=0$, 
(b) $\int_0^t \diff{s_1}\int_0^t \diff{s_2}\int_0^t \diff{s_3} \left\langle \overline{\xi}(t)\overline{\xi}(s_1)\overline{\xi}(s_2)\overline{\xi}(s_3)\right\rangle= 3 \sigma^2 \int_0^t \diff{s_1} K(t-s_1) K(s_1)\diff{s_1}$ for $n=1$ 
and for general $n>1$:
\begin{equation}
\int_0^{t}\diff{s_1}\ldots\int_0^{t}\diff{s_{1+2 n}} 
\left\langle\overline{\xi}(t)\overline{\xi}(s_1)\ldots\overline{\xi}(s_{1+2 n})
\right\rangle = \frac{\sigma^{1+n}(1+2 n)!}{2^n} 
\int_0^{t}\! \diff{s_n}K(t-s_n) \!
\prod_{m=2}^{n}\int_0^{s_m}\! \diff{s_m} K(s_{m}-s_{m-1}) K(s_1).
\label{eq:xiBsuperDI}
\end{equation}   
Substituting these results into Eq.~\eqref{eq:xiBF}, we find ($s_n\!=\!s$): 
\begin{align}
\left\langle \overline{\xi}(t) h(k,t) \right\rangle &=
i k \sigma e^{-i k v_0 t} K(t) + i k \sigma
\times \notag\\ & \qquad \times 
\int_0^t \diff{s} K(t-s) e^{-i k v_0 t}  
\left[ \frac{\sigma (i k)^2}{2} K(s) 
+\sum_{n=2}^{\infty}\frac{\sigma^{n} (i k)^{2 n}}{2^{n}} \prod_{m=2}^{n} \int_0^{s_m}\diff{s_{m-1}}K(s_{m}-s_{m-1}) K(s_1)
\right] . 
\label{eq:xiBNovikovF}
\end{align}
Comparing Eqs.~\eqref{eq:xiBfuncTaylor2}, \eqref{eq:xiBNovikovF}, we obtain the equation:
\begin{align}
\left\langle \overline{\xi}(t) h(k,t) \right\rangle &=
i k \sigma e^{-i k v_0 t} K(t)+ i k \sigma \int_0^t \diff{s} K(t-s) e^{-i k v_0 (t-s)} \left[ i k v_0 + \derpar{}{s}\right]\widetilde{P}(k,s) \notag\\
&= i k \sigma \left[i k v_0 +\derpar{}{t} \right]
\int_0^t \diff{s} K(t-s) e^{-i k v_0 (t-s)} \widetilde{P}(k,s).
\label{eq:xiBNovikov}
\end{align}
The equivalence of the two expressions at the rhs of Eq.~\eqref{eq:xiBNovikov} is proved by taking their Laplace transforms. Substituting this formula into Eq.~\eqref{eq:GaussCorrFP2} yields the Fourier transform of Eq.~\eqref{fracad}.


\section{Special Functions: Definitions and Useful Relations}
\label{specialF}

Here, we review definitions and useful properties of the three parameter Mittag-Leffler function and the Fox H-function. For further details on these special functions and derivations of the relations presented below we refer to \cite{haubold2011mittag}.  

\subsection{The Fox H-Function}
\label{App5.2}

The Fox H-function is formally defined in terms of the following Mellin-Barnes type integral: 
\begin{equation}
H_{p,q}^{m,n}\left[ z
\left|
\begin{array}{l} 
(a_1,A_1),\ldots ,(a_p,A_p) \\ [0.1cm]
(b_1,B_1), \ldots ,(b_q,B_q)
\end{array}
\right.\right]=\frac{1}{2 \pi i}\int_{\Omega} \Theta(s)\,z^{-s}\diff{s},
\label{eq:FHfunction}
\end{equation}
where $i=(-1)^{-1/2}$, $z \neq 0$ and $z^{-s}=\exp{\left[-s\left(\ln |z|+i\arg z\right)\right]}$. Here, $\ln |z|$ stands for the natural logarithm of $|z|$, whereas $\arg z$ is not necessarily its principal value. The function $\Theta(s)$ is defined in terms of Gamma functions as  
\begin{equation}
\Theta(s)=\frac{\left\{\prod_{j=1}^{m}\Gamma(b_j + B_j s)\right\}\left\{\prod_{j=1}^{n}\Gamma(1 - a_j - A_j s)\right\}}{\left\{\prod_{j=1+m}^{q}\Gamma(1 - b_j - B_j s)\right\}\left\{\prod_{j=1+n}^{p}\Gamma(a_j + A_j s)\right\}},
\label{eq:thetaFH}
\end{equation}
where $m, n, p, q \in \mathbb{N}_0$ with $0 \leq n \leq p$ and $1 \leq m \leq q$; $A_i, B_j \in \mathbb{R}_{+}$; $a_i,b_j \in \mathbb{R}$ (or alternatively $\mathbb{C}$) with $i=1,\ldots,p$ and $j=1,\ldots,q$. Any empty product in Eq.~\eqref{eq:thetaFH} is to be interpreted as unity. The contour $\Omega$ in Eq.~\eqref{eq:FHfunction} is suitably chosen to separate the poles $\xi_{j \nu}=-(\nu+b_j)/B_j$, with $j=1,\ldots,m$ and $\nu \in \mathbb{N}_0$, of $\Gamma(b_j + B_j s)$ from the poles $\chi_{i \nu}=(1-a_i+\nu)/A_i$, with $i=1,\ldots,n$ and same $\nu$, of $\Gamma(1 - a_j - A_j s)$. Thus, the condition $A_i (b_j+\nu) \neq B_j (a_i-1-\nu)$ ensures the existence of the contour $\Omega$ and consequently the convergence of the integral in Eq.~\eqref{eq:FHfunction}. 
A popular choice for the contour $\Omega$ consists in a path running parallel to the imaginary axis from $\gamma - i\, \infty$ to $\gamma + i\, \infty$, where $\gamma \in \mathbb{R}=(-\infty,+\infty)$ is chosen arbitrarily such that it separates all the poles $\xi_{j \nu}$ from all the poles $\chi_{i \nu}$. If we choose such a contour, the convergence of the Mellin-Barnes integral in Eq.~\eqref{eq:FHfunction} is obtained if $a^* > 0$ and $|\arg z|< (\pi/2) a^*$, $z \neq 0$, with $a^*$ being the following parameter: 
\begin{equation}
a^*=\sum_{j=1}^{n} A_j - \sum_{j=n+1}^{p} A_j + \sum_{j=1}^{m} B_j - \sum_{j=m+1}^{q} B_j. 
\label{eq:alphaP}
\end{equation}
The integral also converges if $a^*=0$, $\gamma \mu + \operatorname{Re}(\delta)<-1$, $\arg z = 0$ and $z \neq 0$, where
\begin{equation}
\delta=\sum_{j=1}^{q} b_j - \sum_{j=1}^{p} a_j + \frac{p-q}{2}.
\label{eq:A.3}
\end{equation}
Other equivalent choices of $\Omega$, with the corresponding convergence conditions for the integral of Eq.~\eqref{eq:FHfunction}, are available. 
A first useful property of the H-function is its symmetry under exchange of the pairs of parameters $(a_p,A_p)$ and/or $(b_p,B_p)$. Specifically, the H-function is symmetric under permutations of the pairs $(a_i,A_i)$ for $i=1,\ldots,n$ or separately for $i=n+1,\ldots,p$; likewise it is symmetric if we make a permutation of the pairs $(b_j,B_j)$ for $j=m+1,\ldots,q$ or separately for $j=1,\ldots,m$. 
A second property enables us to reduce the order of the function if one of the pairs $(a_i,A_i)$ for $i=1,\ldots,n$ is equal to one of the pairs $(b_j,B_j)$ for $j=1+m,\ldots,q$ or alternatively for $i=1+n,\ldots,p$ and $j=1,\ldots,m$. In these different cases, the H-function reduces to one of lower order with p, q and n (or m respectively) decreased by one. In formulas, we have:  
\begin{equation}
H_{p,q}^{m,n}\!\left[ z
\left|
\begin{array}{l} 
(a_1,A_1),\ldots,(a_p,A_p) \\ [0.1cm]
(b_1,B_1),\ldots,(b_{q-1},B_{q-1}),(a_1,A_1)
\end{array}
\right.\right]
=
H_{p-1,q-1}^{m,n-1}\!\left[ z
\left|
\begin{array}{l} 
(a_2,A_2),\ldots,(a_p,A_p) \\ [0.1cm]
(b_1,B_1),\ldots,(b_{q-1},B_{q-1})
\end{array}
\right.\right],
\label{eq:reduction1}
\end{equation}
provided $n \geq 1$ and $q > m$; and alternatively: 
\begin{equation}
H_{p,q}^{m,n}\!\left[ z
\left|
\begin{array}{l} 
(a_1,A_1),\ldots,(a_{p-1},A_{p-1}),(b_1,B_1) \\ [0.1cm]
(b_1,B_1),\ldots,(b_q,B_q)
\end{array}
\right.\right]
=
H_{p-1,q-1}^{m-1,n}\!\left[ z
\left|
\begin{array}{l} 
(a_1,A_1),\ldots,(a_{p-1},A_{p-1}) \\ [0.1cm]
(b_2,B_2),\ldots,(b_{q},B_{q})
\end{array}
\right.\right],
\label{eq:reduction2}
\end{equation}
provided $m \geq 1$ and $p > n$. 
The Fox H-function satisfies the following scaling relation:
\begin{align}
H_{p,q}^{m,n}\!\left[ z^{r}
\left|
\begin{array}{l} 
(a_p,A_p) \\ [0.1cm]
(b_q,B_q)
\end{array}
\right.\right]
&=\frac{1}{r}
H_{q,p}^{n,m}\!\left[ z
\left|
\begin{array}{l} 
\left(a_p,A_p/r\right) \\ [0.1cm]
\left(b_q,B_q/r\right)
\end{array}
\right.\right] ,
\quad \forall r \in \mathbb{R}_+/\{0\} .
\label{eq:scalingFH}
\end{align}
Two further properties enable us either to invert the independent variable inside the H-function:   
\begin{equation}
H_{p,q}^{m,n}\!\left[ z
\left|
\begin{array}{l} 
(a_p,A_p) \\ [0.1cm]
(b_q,B_q)
\end{array}
\right.\right]
=
H_{q,p}^{n,m}\!\left[ \frac{1}{z}
\left|
\begin{array}{l} 
(1-b_q,B_q) \\ [0.1cm]
(1-a_p,A_p)
\end{array}
\right.\right] 
\label{eq:invertFH}
\end{equation}
or to absorb powers of the independent variable of general exponent $\sigma \in \mathbb{C}$ inside the H-function: 
\begin{equation}
z^{\sigma} H_{p,q}^{m,n}\!\left[ z
\left|
\begin{array}{l} 
(a_p,A_p) \\ [0.1cm]
(b_q,B_q)
\end{array}
\right.\right]
=
H_{p,q}^{m,n}\!\left[ z
\left|
\begin{array}{l} 
(a_p+\sigma A_p,A_p) \\ [0.1cm]
(b_q+\sigma B_q,B_q)
\end{array}
\right.\right].
\label{eq:absorbFH}
\end{equation}
On the one hand, the Mellin-cosine(sine) transform of the Fox H-function is given by \cite{prudnikov1990integrals}: 
\begin{multline}
\int_0^{\infty} z^{\rho-1}
\left\{ \!\!\!
\begin{array}{l}
\sin{(\kappa z)} \\ [0.1cm]
\cos{(\kappa z)}
\end{array}
\!\!\!
\right\}
H_{p,q}^{m,n}\!\left[ a z^{r}
\left|
\begin{array}{l} 
(a_p,A_p) \\ [0.1cm]
(b_q,B_q)
\end{array}
\right.\right]
\diff{z}= 
\frac{2^{\rho - 1}\sqrt{\pi}}{\kappa^{\rho}}
H_{p+2,q}^{m,n+1}\!\left[ a \left(\frac{2}{\kappa}\right)^r
\left|
\begin{array}{l} 
\left(\left(\frac{3\mp1-2 \rho}{4}\right),\frac{r}{2}\right),(a_p,A_p),\left(\left(\frac{3 \pm 1 - 2 \rho}{4}\right),\frac{r}{2}\right) \\ [0.1cm]
(b_q,B_q)
\end{array}
\right.\!\!\right]
\label{eq:FHCStransf}
\end{multline}
where the following conditions must be satisfied: (i) $a^*,r,\kappa>0$, (ii) $|\arg(a)|<a^*\pi/2$, (iii) $\operatorname{Re}{(\rho)}+r \min_{1 \le j \le m}\, \operatorname{Re}{\left(\frac{b_j}{B_j}\right)}> \frac{(-1 \mp 1)}{2}$, (iv) $\operatorname{Re}{(\rho)+r \max_{1 \le j \le n}}\, \operatorname{Re}{\left(\frac{a_j-1}{A_j}\right)}< 1$. 
On the other hand, the Mellin transform of a general H-function is  
\begin{equation}
\int_{0}^{\infty}z^{\xi - 1}
H_{p,q}^{m,n}\!\left[ a z 
\left|
\begin{array}{l} 
(a_p,A_p) \\ [0.1cm]
(b_q,B_q)
\end{array}
\right.\right]
\diff{z}=a^{-\xi} \Theta(\xi) ,
\label{eq:integralFH}
\end{equation}
with $\Theta$ defined as in Eq.~\eqref{eq:thetaFH}.
In conclusion, we provide a formula for the general n-th order derivative of the H-function, i.e.,   
\begin{equation}
\frac{\diff{\!}^{\,r}}{\diff{x^r}}
H_{p,q}^{m,n}\!\left[ (c\,x+d)^h 
\left|
\begin{array}{l} 
(a_p,A_p) \\ [0.1cm]
(b_q,B_q)
\end{array}
\right.\right]
=
\left(\frac{c}{c\,x+d}\right)^r \!
H_{1+p,1+q}^{m,1+n}\!\left[ (c\,x+d)^h 
\left|
\begin{array}{l} 
(0,h), (a_p,A_p) \\ [0.1cm]
(b_q,B_q), (r,h)
\end{array}
\right.\right] .
\label{eq:derivativeFH}
\end{equation}

\subsection{The Three Parameter Mittag-Leffler Function}
\label{Sec5.1}

The three parameter Mittag-Leffler function is defined by the following power-series:
\begin{align}
\gML{\alpha}{\beta}{\delta}{z}=\sum_{n=0}^{\infty}\frac{(\delta)_n}{\Gamma\left(\beta + \alpha\, n\right)}\frac{z^{n}}{n!} ,
\label{eq:gMLDEF}
\end{align}
where $(\delta)_n= \Gamma(\delta + n)/\Gamma(\delta)$ is the Pochhammer symbol. The two and one parameter Mittag-Leffler functions $E_{\alpha,\beta}\left( z \right)$ and $E_{\alpha}\left( z \right)$ are obtained as special cases of Eq.~(G13) 
by setting $\delta=1$, and also $\beta=1$ for the latter one. Its Laplace transform is 
\begin{equation}
\mathcal{L}\left\{z^{\beta\,-\,1}\gML{\alpha}{\beta}{\delta}{\pm c\,z^{\alpha}}\right\}(\lambda)=\frac{\lambda^{\alpha\,\delta-\beta}}{(\lambda^{\alpha}\mp c)^{\delta}} 
\label{eq:Lap3ML}
\end{equation}
with $\operatorname{Re}(\lambda)>|c|^{1/\alpha}$. 
The three parameter Mittag-Leffler function can be expressed as a Fox H-function as 
\begin{equation}
\gML{\alpha}{\beta}{\delta}{\pm\,z}=\frac{1}{\Gamma(\delta)}
H_{1 2}^{1 1}\!\left[\mp\,z
\left| \!\!
\begin{array}{l} 
(1-\delta,1) \\ [-0.01cm]
(0,1),(1-\beta,\alpha)
\end{array}
\!\!\!\right.\right]. 
\label{eq:3MLtoFH}
\end{equation}
This formula is derived by solving the corresponding integral of Eq.~\eqref{eq:FHfunction} with the residue theorem. 
In several anomalous diffusive systems, this function plays a major role, as it typically describes their mean square displacement 
(in this case then $z$ is the time variable). 
It is then important to study its asymptotic scaling for both small and large values of $z$. In the former case, the function $\gML{\alpha}{\beta}{\delta}{-z^{\alpha}}$ behaves as a stretched exponential. In fact, by looking at Eq.~(G13)
, we can write:  
\begin{align}
\gML{\alpha}{\beta}{\delta}{-z^{\alpha}}&\sim \frac{1}{\Gamma(\beta)}-\delta\,\frac{z^{\alpha}}{\Gamma(\alpha + \beta)} 
\sim \frac{1}{\Gamma(\beta)}\exp{\left(-\delta\,\frac{\Gamma(\beta)}{\Gamma(\alpha +\beta)}z^{\alpha}\right)} .
\label{eq:3MLshortS}
\end{align}
In the latter case, it is convenient to look at the equivalent definition (valid for $|z|>1$) \cite{saxena2004unified}   
\begin{equation}
\gML{\alpha}{\beta}{\delta}{-z}=\frac{z^{-\delta}}{\Gamma(\delta)}\sum_{n=0}^{\infty}\frac{\Gamma(\delta + n)}{\Gamma\left(\beta - \alpha (\delta + n)\right)}\frac{z^{-n}}{n!},
\label{eq:gMLDEF2}
\end{equation}
which then predicts a asymptotic power-law behaviour for $|z| \gg 1$, i.e.,  
\begin{equation}
\gML{\alpha}{\beta}{\delta}{-z^{\alpha}}\sim \frac{z^{-\alpha\,\delta}}{\Gamma(\beta -\alpha\,\delta)} .
\label{eq:3MLlongS}
\end{equation}

\twocolumngrid


\end{document}